\documentclass[twocolumn]{aastex62}

\def\chandra{{\itshape Chandra\/}}

\def\spitzer{{\itshape Spitzer\/}}

\def\wise{{\itshape WISE\/}}
\def\galex{{\itshape GALEX\/}}

\def\xmm{{\itshape XMM-Newton\/}}
\def\nustar{{\itshape NuSTAR\/}}

\def\xray{\hbox{X-ray}}

\def\etal{{et\,al.}}

\def\lxsfr{$L_{\rm X}$/SFR}
\def\lgoh{$12 + \log({\rm O}/{\rm H})$}
\def\ltsima{$\; \buildrel < \over \sim \;$}
\def\simlt{\lower.5ex\hbox{\ltsima}}
\def\gtsima{$\; \buildrel > \over \sim \;$}
\def\simgt{\lower.5ex\hbox{\gtsima}}
\def\kms{\ifmmode{~{\rm km~s^{-1}}}\else{~km s$^{-1}$}\fi}
\def\lsim{\lower0.3em\hbox{$\,\buildrel <\over\sim\,$}}
\def\gsim{\lower0.3em\hbox{$\,\buildrel >\over\sim\,$}}

\def\h2{H$_2$}
\def\flux{erg~cm$^{-2}$~s$^{-1}$}
\def\lum{erg~s$^{-1}$}

\def\sfr{$M_{\odot}$~yr$^{-1}$}

\def\aap{A\&A}
\def\apj{ApJ}
\def\apjl{ApJL}
\def\apjs{ApJS}
\def\aj{AJ}
\def\mnras{MNRAS}
\def\araa{ARA\&A}

\def\nat{Nature}

\def\ngal{33}  
\def\nsup{22}  
\def\nobs{137}
\def\ntot{55}
\def\nxps{1311}
\def\nhmxb{49}
\def\nlmxb{22}
\def\ncxb{29}

\usepackage{underscore}
\usepackage{longtable}
\usepackage{amsmath}
\usepackage{apjfonts}
\shorttitle{Metallicity-Dependent HMXB Luminosity Functions}
\shortauthors{Lehmer et al.}

\begin{document}

%
\title{{\bf {\large The Metallicity Dependence of the High-Mass X-ray Binary Luminosity Function}}}

\correspondingauthor{Bret Lehmer}
\email{lehmer@uark.edu}

\author[0000-0003-2192-3296]{Bret~D.~Lehmer}
\affil{Department of Physics, University of Arkansas, 226 Physics Building, 825 West Dickson Street, Fayetteville, AR 72701, USA}

\author[0000-0002-2987-1796]{Rafael~T.~Eufrasio}
\affil{Department of Physics, University of Arkansas, 226 Physics Building, 825 West Dickson Street, Fayetteville, AR 72701, USA}

\author{Antara Basu-Zych}
\affiliation{NASA Goddard Space Flight Center, Code 662, Greenbelt, MD 20771, USA}
\affiliation{Center for Space Science and Technology, University of
Maryland Baltimore County, 1000 Hilltop Circle, Baltimore, MD 21250, USA}

\author{Keith~Doore}
\affil{Department of Physics, University of Arkansas, 226 Physics Building, 825 West Dickson Street, Fayetteville, AR 72701, USA}

\author[0000-0003-1474-1523]{Tassos Fragos}
\affiliation{Geneva Observatory, 
University of Geneva, 
Chemin des Maillettes 51, 
1290 Sauverny, Switzerland}

\author{Kristen Garofali}
\affil{Department of Physics, University of Arkansas, 226 Physics Building, 825 West Dickson Street, Fayetteville, AR 72701, USA}
\affiliation{NASA Goddard Space Flight Center, Code 662, Greenbelt, MD 20771, USA}

\author{Konstantinos Kovlakas}
\affil{Physics Department, University of Crete, GR 71003, Heraklion, Greece}
\affil{Institute of Astrophysics, Foundation for Research and Technology-Hellas, GR 71110 Heraklion, Greece}

\author{Benjamin F. Williams}
\affil{Department of Astronomy, Box 351580, University of Washington, Seattle, WA 98195, USA}

\author{Andreas Zezas}
\affiliation{Harvard-Smithsonian Center for Astrophysics, 60 Garden Street, Cambridge, MA 02138, USA}
\affiliation{Foundation for Research and Technology-Hellas, 100 Nikolaou Plastira Street, 71110 Heraklion, Crete, Greece}
\affiliation{Physics Department \& Institute of Theoretical \& Computational Physics, P.O. Box 2208, 71003 Heraklion, Crete, Greece}

\author{Luidhy Santana-Silva}
\affiliation{NAT-Universidade Cruzeiro do Sul / Universidade Cidade de S{\u a}o Paulo, Rua Galv{\u a}o Bueno, 868, 01506-000, S{\u a}o Paulo, SP, Brazil}

%
\begin{abstract}
%

We present detailed constraints on the metallicity dependence of the high mass
X-ray binary (HMXB) \xray\ luminosity function (XLF).  We analyze $\approx$5~Ms
of \chandra\ data for \ntot\ actively star-forming galaxies at $D \simlt
30$~Mpc with gas-phase metallicities spanning \lgoh~$\approx$~7--9.2.  Within
the galactic footprints, our sample contains a total of \nxps\ \xray\ point sources, of
which $\approx$\nhmxb\% are expected to be HMXBs, with the remaining sources
likely to be low-mass \xray\ binaries (LMXBs; $\approx$\nlmxb\%) and unrelated
background sources ($\approx$\ncxb\%).  We construct a model that successfully
characterizes the average HMXB XLF over the full metallicity range.  We
demonstrate that the SFR-normalized HMXB XLF shows clear trends with
metallicity, with steadily increasing numbers of luminous and ultraluminous
\xray\ sources ($\log L$(\lum)~=~38--40.5) with declining metallicity.
However, we find that the low-luminosity ($\log L$(\lum)~=~36--38) HMXB XLF
appears to show a nearly constant SFR scaling and slope with metallicity.  Our
model provides a revised scaling relation of integrated \lxsfr\ versus \lgoh\ and
a new characterization of its the SFR-dependent stochastic scatter.  The general trend of
this relation is broadly consistent with past studies based on integrated
galaxy emission; however, our model suggests that this relation is driven primarily 
by the high-luminosity end of the HMXB XLF.  Our results have implications for
binary population synthesis models, the nature of super-Eddington accreting
objects (e.g., ultraluminous X-ray sources), recent efforts to identify active
galactic nucleus candidates in dwarf galaxies, and the \xray\ radiation fields
in the early Universe during the epoch of cosmic heating at $z \simgt 10$.

%
\end{abstract}
%

\keywords{stars: formation --- galaxies: normal --- X-rays: binaries --- X-rays: galaxies }

%
\section{Introduction}
%

Over the last $\approx$20 years, the {\it Chandra \xray\ Observatory} ({\it
Chandra}) has accumulated an extensive archive of \xray\ imaging data capable
of resolving populations of \xray\ binaries (XRBs) within relatively nearby
galaxies ($\simlt$100~Mpc).  Thanks to these \chandra\ data, along with
complementary multiwavelength data (e.g., from \galex, \spitzer, \wise, and
SDSS), it is now possible to conduct detailed statistically meaningful studies
that characterize how the XRB \xray\ luminosity functions (XLFs) vary with
local physical properties (e.g., star-formation rate (SFR) and stellar mass
($M_\star$)).  
Such XRB XLF scaling relations provide important benchmarks for understanding
\xray\ emission from a variety of galaxy populations, compact object formation
and evolution, the binary phase of stellar evolution, and several
close-binary subclasses (e.g., gravitational-wave sources, short gamma-ray bursts,
and millisecond pulsars).  

On the observational side, XRB scaling relations are routinely used in
assessing whether \xray\ detected sources are consistent with stellar-mass XRBs
or accreting supermassive black holes (SMBHs; see Mezcua \etal\ 2018, Latimer
\etal\ 2019, Birchall \etal\ 2020, Hodges-Kluck \etal\ 2020, Koudmani \etal\
2020, and Secrest \etal\ 2020 for recent examples).  These assessments are
often based on XLF-integrated scaling relations and their galaxy-population
statistical scatters (e.g., the $L_{\rm X}$-SFR relation; Ranalli \etal\ 2003;
Colbert \etal\ 2004; Persic \& Rephaeli~2007; Lehmer \etal\ 2008, 2010, 2016,
2019, 2020; Mineo \etal\ 2012a, 2012b; Basu-Zych \etal\ 2013a; Aird \etal\
2017).  However, due to the often poor sampling of the bright-source ends of
XLFs, these relations are likely to become non-linear with highly-skewed
scatter distributions in the low-SFR and low-$M_\star$ regimes where they are
often applied (e.g., Gilfanov \etal\ 2004; Justham \& Schawinski~2012; Lehmer
\etal\ 2019).  A detailed characterization of how the XRB XLF varies with
galaxy physical properties can provide appropriate statistical quantification
of how \xray\ scaling relations and their scatter vary across these properties.

Thus far, most XRB XLF scaling relation studies have focused on constraining
the high-mass XRB (HMXB) and low-mass XRB (LMXB) XLF scalings with SFR and
$M_\star$, respectively (e.g., Grimm \etal\ 2002; Gilfanov~2004; Mineo
\etal\ 2012a; Zhang \etal\ 2012; Lehmer \etal\ 2019, 2020).  However, recently
it has become clear, from both theoretical and observational studies of
integrated scaling relations, that additional physical parameters, including
star-formation history (SFH) and metallicity, have important impacts on these
scaling relations (e.g., Dray~2006; Mapelli \etal\ 2010; Kaaret \etal\ 2011;
Basu-Zych \etal\ 2013b, 2016; Fragos \etal\ 2013a,b; Brorby \etal\ 2014, 2016;
Douna \etal\ 2015; Lehmer \etal\ 2017; Antoniou \etal\ 2019; Fornasini \etal\
2019, 2020; Kouroumpatzakis \etal\ 2020; Kovlakas \etal\ 2020).  

Particular attention has been directed to the metallicity dependence of the
HMXB XLF. HMXBs are the most luminous XRB population per stellar mass, and are
expected to provide important ionizing radiation from stellar populations that
both lasts longer than the far-UV emission from young massive stars and
traverses longer path lengths before being absorbed.  Given these properties,
HMXBs have been proposed to play roles in (1) interstellar medium feedback (e.g.,
Pakull \etal\ 2010; Soria \etal\ 2010, 2014; Justham \& Schawinski~2012;
L{\'o}pez \etal\ 2019); (2) the production of Lyman continuum radiation, as well as He~II
and Lyman-$\alpha$ emission lines from starbursts (e.g., Kaaret \etal\ 2017a;
Bleum \etal\ 2019; Schaerer \etal\ 2019; Svoboda \etal\ 2019; Dittenber \etal\
2020; Saxena \etal\ 2020; Senchyna \etal\ 2020); and (3) the heating of the early
intergalactic medium at $z \simgt$~10 (e.g., Mirabel \etal\ 2011; Mesinger
\etal\ 2013; Pacucci \etal\ 2014; Madau \& Fragos~2017; Das \etal\ 2017; Greig
\& Mesinger~2018; Park \etal\ 2018; Heneka \& Mesinger~2020).

While some basic models have been proposed for the HMXB XLF dependendence on
metallicity (e.g., Brorby \etal\ 2014; Basu-Zych \etal\ 2016; Ponnada \etal\
2020), these models are primarily based on interpreting the $L_{\rm X}$/SFR
versus metallicity relation (hereafter, the $L_{\rm X}$--SFR--$Z$ relation) in
terms of a varying HMXB XLF normalization and power-law slope.  Furthermore,
studies that place direct constraints on the metallicity-dependence of
point-source populations within galaxies have focused primarily on the
ultraluminous X-ray source (ULX; $L > 10^{39}$~\lum) formation rate per unit
SFR (e.g., Zampieri \& Roberts~2009; Mapelli \etal\ 2010; Prestwich \etal\
2013; Douna \etal\ 2015; Kovlakas \etal\ 2020).  These studies find a clear
decline in the ULX rate with increasing metallicity; however, they do not
characterize the luminosity dependence of this trend nor do they make a direct
connection with more typical lower-luminosity XRB populations.  

In a recent study of the XLFs of 38 nearby ($D < 30$~Mpc) galaxies (primarily
late-type galaxies), Lehmer \etal\ (2019; hereafter, L19), we showed that the five
lowest metallicity galaxies in the sample hosted a factor of
$\approx$2--10 times excess of HMXBs with $L \simgt 5 \times 10^{38}$~\lum\
compared to the average galaxy population, while the number of $L \approx$~(0.1--5)~$\times
10^{38}$~\lum\ sources in these relatively low-metallicity galaxies appeared to
be consistent with the full galaxy sample.  This suggests that the
metallicity-dependent XLF evolution is more complex than a simple scaling in
the overall normalization and power-law slope of the XLF.

In this paper, we expand on the early results from L19 and more fully characterize
how the SFR-scaled HMXB XLF normalization and detailed shape vary with metallicity. 
In Section~2, we discuss our galaxy sample selection and present the galaxy
properties.  In Section~3, we briefly summarize our \chandra\ data analysis
procedure, which follows closely the approach outlined in L19.  In Section~4,
we present our global metallicity-dependent HMXB XLF model and fitting results.
In Section~5, we discuss our results in the context of past studies and
highlight problems to address in future studies.  Finally, Section~6 provides a summary of the key results from this work.

Unless noted otherwise, we make use of multiwavelength fluxes and luminosities
that have been corrected for Galactic absorption, but not intrinsic absorption.
Throughout, we quote \xray\ luminosities in the \hbox{0.5--8~keV} bandpass and
use $L$ to denote individual point-source luminosities and $L_{\rm X}$ to
represent population-integrated luminosities.  For clearer comparisons with
past studies, we often quote SFR and $M_\star$ values that are based on a
Kroupa~(2001) initial mass function (IMF) at solar abundances, with a SFH that
is constant over 13.6~Gyr.  As we discuss in Section~2, the assumption of a
constant abundance for these calculations has only a minor impact on our
results.

%
\section{Sample Selection and Properties}
%

Given our goal to study the effects of metallicity on HMXB formation, we
constructed relatively nearby ($D < 30$~Mpc) samples of star-forming galaxies
with high specific SFRs (sSFR~$\equiv$~SFR/$M_\star \simgt 10^{-10}$~yr$^{-1}$)
that span a broad range of metallicity (\lgoh~$\approx$~7--9.2; $Z \approx
0.02$--3~$Z_\odot$).\footnote{Throughout this paper we quote metallicities in
terms of gas-phase oxygen abundances, \lgoh, and take the solar value to be
8.69 (Allende~Prieto \etal\ 2001).  When comparing with theoretical models that
specify metallicities explicitly, we assume the conversion $\log(Z/Z_\odot) =
\log({\rm O}/{\rm H})-\log({\rm O}/{\rm H})_\odot = \log({\rm O}/{\rm
H})-3.31$.}  The adopted sSFR boundary limits contributions from LMXBs
that can dominate in lower sSFR galaxies (e.g., Lehmer \etal\ 2010;
Kouroumpatzakis \etal\ 2020).  We first assembled a {\it main sample} of \ngal\
such objects that had both \chandra\ archival data from the literature and a
suite of multiwavelength observations spanning the UV to mid-IR.  Our main
sample contains (1) 23 galaxies from the spectroscopic study from Moustakas
\etal\ (2010) of galaxies in the {\it Spitzer} Infrared Nearby Galaxies Survey
(SINGS; Kennicutt \etal\ 2003); (2) six objects from Engelbracht \etal\ (2008);
and (3) four additional nearby galaxies (NGC~5253, M83, M101, and DDO68) with
deep \chandra\ exposures ($>$100~ks).  We chose not to include well-studied
disk-dominated galaxies with
high inclination angles to our line of sight ($\simgt$70~deg; e.g., M82 and
NGC~253), due to highly varied (and occasionally very large) absorption of
\xray\ point-sources through the galactic disks.  

In addition to our main sample, we assembled a {\it supplemental sample} of \nsup\
extremely metal poor objects with \lgoh~$< 8$ and $D<30$~Mpc that were
drawn from Prestwich \etal\ (2013).  We note that these galaxies have been
utilized in the Brorby \etal\ (2014) and Douna \etal\ (2015) studies of \xray\
properties of extremely metal-poor galaxies and include many of the lowest
metallicity galaxies known in the local Universe.  Galaxies in the
supplementary sample are compact ($\approx$0\farcm4--1\farcm4 semi-major axes)
dwarf galaxies, and have relatively shallow \chandra\ and multiwavelength data
compared to our main sample.  Therefore, physical parameters are more uncertain
for this population and \xray\ constraints are limited to only bright point
sources in the galaxies, with some uncertainty on whether they are XRB or SMBH
dominated.

All galaxies in our samples have oxygen abundance measurements available from
either strong-line calibrations or a ``direct,'' electron-temperature-based
theoretical calibration.  The strong-line estimates were originally based on
either the theoretical Kobulnicky \& Kewley (2004; herafter, KK04) or empirical
Pettini \& Pagel (2004; PP04) calibrations.  KK04 estimates were available for
the 23 SINGS galaxies in our main sample via Moustakas \etal\ (2010), who
utilized the $R_{23}$ emission-line ratio, $R_{23} =$~([O~{\small
II}]$\lambda$3727 + [O~{\small III}] $\lambda \lambda$4959, 5007)/H$\beta$, and
[O~{\small III}]$\lambda$5007/[O~{\small II}]$\lambda$3727.  PP04 estimates
were available for M83 and M101 (Bresolin~\etal\ 2009 and Hu \etal\ 2018), and
are based on the ratio ([O~{\small III}]$\lambda$5007/H$\beta$)/([N~{\small
II}]$\lambda$6584/H$\alpha$).  All other galaxies, including eight main-sample
galaxies (\lgoh~$\simlt 8.4$), as well as the \nsup\ extremely metal-poor
supplemental-sample galaxies, had estimates based on the direct method (see
Table~1 for references), which uses the [O~{\small III}]$\lambda\lambda$4959,
5003/[O~{\small III}]$\lambda$4363 ratio. 

\begin{deluxetable*}{lcccccccccrcr}
\tablewidth{1.0\columnwidth}
\tabletypesize{\footnotesize}
\tablecaption{HMXB-Dominant Galaxy Sample and Properties}
\tablehead{
\multicolumn{1}{c}{}  &  \colhead{} &\colhead{} & \colhead{} & \multicolumn{3}{c}{\sc Size Parameters} & \colhead{} & \colhead{} & \colhead{} & \colhead{} & \colhead{} & \colhead{}\\
\vspace{-0.25in} \\
\multicolumn{1}{c}{\sc Galaxy} &  \multicolumn{2}{c}{\sc Central Position} & \colhead{$D$} & \colhead{$a$} & \colhead{$b$} & \colhead{PA} & \colhead{$\log M_\star$} &  \colhead{SFR} & \colhead{$12 + \log({\rm O/H})$} & \colhead{$t_{\rm exp}$} & \colhead{$\log L_{50}$} & \colhead{} \\ 
\vspace{-0.25in} \\
\multicolumn{1}{c}{\sc Name} &  \colhead{$\alpha_{\rm J2000}$} & \colhead{$\delta_{\rm J2000}$} & \colhead{(Mpc)} & \multicolumn{2}{c}{(arcmin)} & \colhead{(deg)} &  \colhead{($M_\odot$)}   & \colhead{(\sfr)} &  \colhead{(dex)} & \colhead{(ks)} & \colhead{(\lum)} & \multicolumn{1}{c}{$N_{\rm X}$}\\ 
\vspace{-0.25in} \\
\multicolumn{1}{c}{(1)} & \multicolumn{1}{c}{(2)} & \multicolumn{1}{c}{(3)} & \colhead{(4)} & \colhead{(5)} & \colhead{(6)} & \colhead{(7)} & \colhead{(8)} & \colhead{(9)} & \colhead{(10)} & \colhead{(11)} & \colhead{(12)} & \colhead{(13)} 
}
\startdata
\multicolumn{13}{c}{Main Sample} \\
\hline 
             NGC0024\dotfill &     00 09 56.5 & $-$24 57 47.3 &    7.30$\pm$2.30 &   1.38 &   0.39 &       43.5 &                      8.64 &                      0.04 &             8.59$\pm$0.18 &  43 & 36.8 &    8 \\
             NGC0337\dotfill &     00 59 50.1 & $-$07 34 40.7 &   22.40$\pm$2.30 &   0.87 &   0.49 &      157.5 &                      9.32 &                      1.09 &             8.44$\pm$0.07 &   9 & 38.5 &    6 \\
       NGC0628 (M74)\dotfill &       01 36 41.8 & +15 47 00.5 &    7.30$\pm$1.40 &   2.10 &   1.80 &       87.5 &                      9.48 &                      0.33 &             8.54$\pm$0.15 & 268 & 36.4 &   43 \\
             NGC0925\dotfill &       02 27 16.9 & +33 34 44.0 &    9.12$\pm$0.17 &   1.87 &   0.82 &      105.0 &                      9.03 &                      0.18 &             8.38$\pm$0.15 &  12 & 37.5 &    7 \\
             NGC1482\dotfill &     03 54 38.9 & $-$20 30 08.8 &   23.20$\pm$2.30 &   1.42 &   0.82 &      105.5 &                     10.43 &                      4.80 &             8.52$\pm$0.16 &  28 & 38.3 &    7 \\
\hline 
\multicolumn{13}{c}{Supplemental Sample}\\
\hline 
  6dFJ0405204-364859\dotfill &     04 05 20.3 & $-$36 49 01.0 &   11.00$\pm$0.00 &   0.23 &   0.19 &       16.5 &                    \ldots &                     0.006 &             7.34$\pm$0.09 &   5 & 38.0 &    0 \\
         HS0822+3542\dotfill &       08 25 55.5 & +35 32 32.0 &   12.70$\pm$0.00 &   0.10 &   0.05 &      141.4 &                    \ldots &                     0.003 &             7.44$\pm$0.06 &   5 & 38.2 &    0 \\
         HS1442+4250\dotfill &       14 44 12.8 & +42 37 44.0 &   10.50$\pm$0.00 &   0.44 &   0.10 &       63.2 &                    \ldots &                     0.011 &             7.64$\pm$0.02 &   5 & 37.8 &    0 \\
               IZw18\dotfill &       09 34 02.0 & +55 14 28.0 &   17.10$\pm$0.00 &   0.13 &   0.09 &      145.0 &                    \ldots &                     0.045 &             7.18$\pm$0.03 &  40 & 37.6 &    1 \\
 J081239.52+483645.3\dotfill &       08 12 39.5 & +48 36 45.0 &    9.04$\pm$0.00 &   0.21 &   0.11 &       73.7 &                    \ldots &                     0.001 &             7.28$\pm$0.06 &   4 & 37.9 &    0 \\
\enddata
\tablecomments{The full version of this table contains 15 columns of information for all \ngal\ and \nsup\ galaxies from our main and supplemental samples, respectively (see Section~2).  An abbreviated version of the table is displayed here to illustrate its form and content.  Col.(1): Adopted galaxy designation with Messier designation, if applicable. Col.(2) and (3): Right ascension and declination of the galactic center. Col.(4): Adopted distance and 1$\sigma$ uncertainties in units of Mpc.  Col.(5)--(7): Isophotal ellipse parameters, including, respectively, semi-major axis, $a$, semi-minor axis, $b$, and position angle east from north, PA.  The majority of the main-sample galaxies have size parameters based on 2MASS $K_s$-band data from Jarrett \etal\ (2003).  In these cases, the $K_s$-band 20~mag~arcsec$^{-2}$ ellipses were used to avoid large contamination from background sources; however, for galaxies with $K_s$-band 20~mag~arcsec$^{-2}$ semi-major axes $a <1$~arcmin, we chose to adopt the ``total' $K_s$-band ellipses.  For the main-sample galaxies DDO68, NGC~4861, and IC2574, and all galaxies in the supplemental sample, we adopted the positions and $D_{25}$ sizes using the HyperLeda database (http://leda.univ-lyon1.fr/).  Col.(8): Logarithm of the galactic stellar mass, $M_\star$, within the regions defined.  Col.(9): Star-formation rate within the defined regions. Col.(10): Adopted estimate of the average oxygen abundances, 12+$\log ({\rm O/H})$; see Col.(15) for references.  For consistency with other studies of XRB scaling relations that include metallicity, we have converted all abundances to the Pettini \& Pagel~(2004; PP04) calibration following the prescriptions in Kewley \& Ellison~(2008). Col.(11): Cumulative \chandra\ exposure time in ks. Col.(12): Logarithm of the 0.5--8~keV luminosity corresponding to 50\% completeness across the galaxy. Col.(13): Number of \xray\ detected sources within the galactic footprints defined by Col.(2)--(3) and (5)--(7). Col.(14) and (15): References to the adopted distances and \lgoh, respectively.\\
References.-- 1=Moustakas \etal\ (2010); 2=Engelbracht \etal\ (2008); 3=Sacchi \etal\ (2016); 4=McQuinn \etal\ (2016); 5=Tully \etal\ (2013); 6=Freedman \etal\ (2001); 7=Nataf \etal\ (2015); 8=Skillman \etal\ (2013); 9=Bresolin \etal\ (2009); 10=Monreal-Ibero \etal\ (2012); 11=Hu \etal\ (2018); 12=Prestwich \etal\ (2013); 13=Guseva \etal\ (2007); 14=Izotov \etal\ (2006); 15=Izotov \& Thuan~(2007); 16=Izotov \etal\ (2007); 17=Izotov \etal\ (2012); 18=Papaderos \etal\ (2008); 19=Pustilnik \etal\ (2003); 20=van Zee~(2000)
}
\end{deluxetable*}

%
%
\begin{figure}
\figurenum{1}
\centerline{
\includegraphics[width=8.8cm]{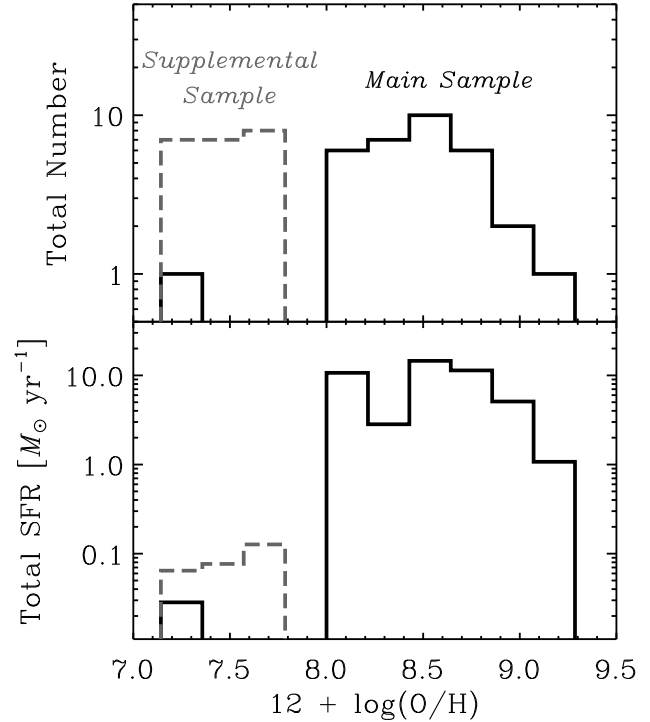}
}
\caption{
Sample distributions of metallicities (\lgoh) plotted in terms of total numbers
of galaxies ({\it top panel\/}) and cumulative SFRs ({\it bottom panel\/}) per
metallicity bin.  Our main sample is shown as {\it solid black histograms} and
our supplemental sample of \nsup\ \lgoh~$< 8$ objects from Prestwich \etal\
(2013) is shown as {\it dashed gray histograms}.  The cumulative SFR is a
proxy for the number of HMXBs that are expected to be present within the galaxy
sample.  The supplemental sample has a low collective SFR and generally shallow
\chandra\ observations.  Although it helps to extend the metallicity range into
the \lgoh~$< 8$ regime, the supplemental sample provides only weak constraints
on the HMXB XLF.
}
\end{figure}

The PP04 strong-line calibrations were empirically calibrated using
direct-method measurements, and the two methods are consistent for
\lgoh~$\approx$~8.09--8.5.  The direct method is generally considered to be
robust for relatively low metallicities (\lgoh~$\simlt 8.5$; e.g.,
Stasi{\'n}ska~2005).  To bring the metallicities for our full galaxy sample
into calibration alignment, we converted all 23 galaxies with KK04 estimates to
the PP04 calibration using the methods described in Kewley \& Ellison~(2008).
In Col.(10) of Table~1, we provide the adopted metallicity values for our
sample galaxies.

Following the procedures in L19, we generated SFR and $M_\star$ maps for
main-sample galaxies using the scaling relations from Hao \etal\ (2011) and
Zibetti \etal\ (2009), respectively.  Values of the SFR and $M_\star$ quoted
for each galaxy were extracted from the maps using the regions defined in
Table~1, with NGC~0628, NGC~5194, and NGC~7552 having central regions removed
due to potential AGN or confusion.  The removal regions were located at the
central positions of the galaxies and were circular in shape with radii of
$r_{\rm remove} =$~3~arcsec for NGC~0628 and NGC~5194, and $r_{\rm remove}
=$~15~arcsec for NGC~7552.

For the supplemental sample galaxies, we utilized the extinction-corrected
values of $L_{\rm FUV}^{\rm corr}$ from Prestwich \etal\ (2013), combined with
the relation of Hao \etal\ (2011), to estimate SFR values.  As per previous
investigations of this sample (Prestwich \etal\ 2013; Brorby \etal\ 2014, 2016;
Douna \etal\ 2015), we do not provide stellar mass estimates for these
galaxies.  These galaxies were originally selected as blue and actively star
forming, and are expected to have high sSFRs and contain negligible
contributions from LMXBs.

As discussed in Section~1, for consistency with other work, the values of SFR
and $M_\star$ above assume conversion factors using solar abundances and do not
incorporate variations due to metallicity.  Applying such a variation would have its
largest impact on the young massive star population, in which low-metallicity
stars are expected to be hotter and produce more UV emission per unit SFR
compared to the solar abundance case (e.g., Bicker \& Fritze-v.~Alvensleben~2005).  We have calculated, using {\ttfamily P\'EGASE} models (Fioc
\& Rocca-Volmerange~1997) for constant SFHs over 100~Myr, that the SFR/$L_{\rm
FUV}$ conversion factors would be $\approx$0.8 and $\approx$1.2 times the
solar-value factor at \lgoh~$=$~7 and 9.2, respectively.  Given that past
results, as well as the results presented below, show a decline in $L_{\rm
X}$/SFR with increasing metallicity, including the metallicity dependence in
the SFR calculations would act to further promote such a trend at a maximium
level of $\approx$0.18~dex across the full metallicity range.

In Table~1, we list the physical properties of galaxies in our main and
supplemental samples.  Our main sample contains galaxies with
SFR~=~0.02--9.0~\sfr\ ($\langle$SFR$\rangle$ = 1.4~\sfr) and $\log
M_\star/M_\odot =$~7.3--10.4 ($\log \langle M_\star/M_\odot \rangle = 9.7$).
Our supplemental sample has a range of SFR~=~0.0007--0.045~\sfr\
($\langle$SFR$\rangle$ = 0.012~\sfr).
Figure~1 shows binned distributions of oxygen abundances for our main and
supplemental samples in terms of numbers of galaxies and cumulative SFR in each
bin.  Given the HMXB XLF scaling with SFR, the SFR of a given bin can be
considered a proxy for the total number of HMXBs present in the samples.  It is
clear that the extremely metal-poor galaxy subsample is expected to harbor
fewer HMXBs than the main sample, given its low cumulative SFR.  

In addition to the relatively low SFRs compared to the main sample, the
supplemental sample has available only relatively shallow \chandra\ exposures
(see Section~3 for a detailed discussion of the \chandra\ data).  The
supplemental sample mean \chandra\ exposure is $\approx$10~ks, compared with
$\approx$150~ks for the main sample.  The combination of these two factors
(relatively low SFR and shallow \chandra\ data) lead to the supplemental sample
containing $\approx$200 times fewer \xray\ detected point sources within the
galaxy footprints compared to the main sample.  In fact, the majority of the
supplemental sample galaxies (17 out of \nsup) contain no detected point
sources coincident with the galaxies, while all main sample galaxies have at
least one detected source.

%
\section{Data Analysis}
%

We chose to incorporate all ACIS-S or ACIS-I \chandra\ imaging observations
with aim-points less than 5~arcmin from the galactic centers (see Col.(3) and
(4) in Table~1).  This led to the selection of \nobs\ and \nsup\ ObsIDs for the
\ngal\ and \nsup\ galaxies in our main and supplementary samples, respectively.
The cumulative \chandra\ exposures spanned the range of 5--853~ks (see Col.(12)
of Table~1).

Our \chandra\ data analysis procedure follows that presented in Section~3.2 of
L19; we briefly describe our procedure here.  For a given galaxy, we
reprocessed all ObsIDs to the most recent calibrations, removed bad events from
bad columns and pixels, and generated exposure maps.  When more than one ObsID
was available, we aligned all observations to the longest exposure, and
co-added all observations for the purpose of identifying \xray\ point sources.
\xray\ point sources were identified by running {\ttfamily wavdetect} on
ObsID-merged images in the 0.5--2~keV, 2--7~keV, and 0.5--7~keV bands, and
resulting point source properties were analyzed using {\ttfamily ACIS Extract}
({\ttfamily AE}; Broos \etal\ 2010, 2012).  For each galaxy, {\ttfamily AE} first
analyzes all ObsIDs separately and subsequently combines them to provide
appropriate source photometry and spectra.  {\ttfamily AE} then performs
spectral fitting and provides flux measurements for all point sources, which
are used for our \xray\ point source catalogs.\footnote{Point-source catalogs
are summarized in Appendix~A and provided in the electronic version of the
paper.}  

%
%
\begin{figure}
\figurenum{2}
\centerline{
\includegraphics[width=9cm]{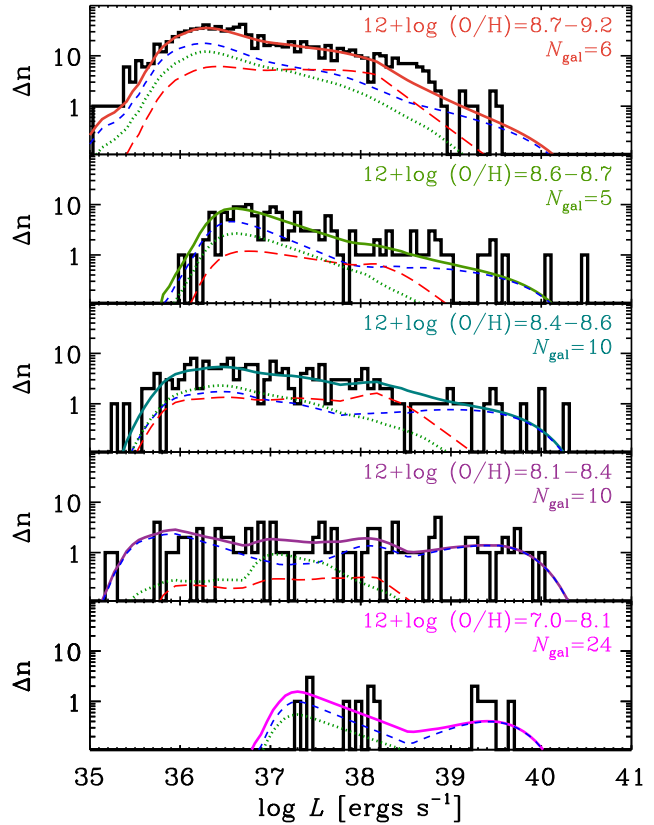}
}
\caption{
Observed distributions of \xray\ point-source luminosities for galaxy subsamples binned
by \lgoh\ ({\it black histograms\/}).  The metallicity range and number of
galaxies within the bin, $N_{\rm gal}$, are annotated in each panel.  The
detailed shapes and \xray\ luminosity ranges of the distributions are dependent
on both the intrinsic distributions of the point-source populations as well as
the completeness characteristics of the observations, which can vary
substantially among galaxies.  Our best-fit metallicity-dependent HMXB XLF
model predictions are shown as {\it short-dashed blue curves}, with the model
contributions from CXB sources and LMXBs shown as {\it dotted green} and {\it
long-dashed red curves}, respectively.  The CXB and LMXB model
contributions for each galaxy vary with \xray\ completeness and scale only with
the galaxy sky coverage and stellar mass, respectively, and do not vary with
metallicity (see Section~3 for details).  The summed contributions of these
model components constitute our total point-source models, which are shown in
each panel as {\it solid curves of varying color}.
}
\end{figure}

%
%
\begin{figure*}[t!]
\figurenum{3}
\centerline{
\includegraphics[width=18cm]{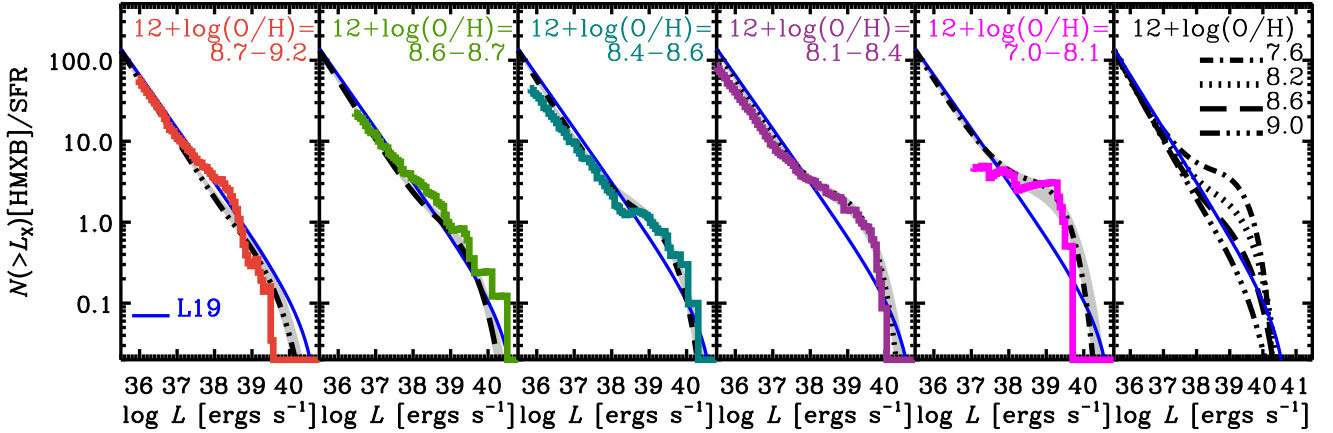}
}
\caption{
{\it First Five Panels}: SFR-normalized cumulative HMXB XLFs for the five
metallicity bins explored in this study.  Data have been plotted in various
colors, following the colors in Figure~2, and are shown here with model CXB and
LMXB contributions subtracted and completeness corrected.  Thus, some regions
in the cumulative plot appear to show small decreases in $N(>L)$ with
decreasing $L$ due to CXB contributions being non-negligible (e.g., in the
low-metallicity regime).  We limited the dispayed luminosity range to be above
the lowest 50\% completeness limit among the galaxies that make up each
subsample.  In each panel, our metallicity-dependent HMXB XLF model ({\it
black curves\/}) is dispayed with 1$\sigma$ uncertainty ({\it gray shaded
region\/}), and the L19 best-fit HMXB model for local galaxies is shown in all
panels as a blue solid line.
{\it Final Panel}: Our best-fit metallicity-dependent HMXB XLF model at various
metallicities (see annotations).  The best-fit model and data reveal a
flattening of the HMXB XLF at $L \simgt 10^{38}$~\lum\ and overall elevation of
the number of bright source per unit SFR with declining metallicity.
}
\end{figure*}

For each galaxy, we assessed point-source completeness (i.e., fraction of
sources recovered by our procedure as a function of luminosity $L$) by running
simulations where mock sources were added to images (using {\ttfamily MARX}
point-spread functions generated by {\ttfamily AE} and custom software) and tested for
recovery by re-running our procedure.  These completeness curves are critical for
correctly modeling the XLFs.  The resulting 50\% completeness limits span $\log
L_{50}$(\lum)~$=$~35.9--38.5 and are listed in Col.(13) of Table~1.  Given the
variations in \chandra\ exposure, distance, and galaxy type, the number of
\xray\ point-sources detected in the galaxy footprints ranges from 0--363 (see
Col.(14) in Table~1).  

To build our metallicity-dependent HMXB XLF model, we first divided our
combined main and supplemental galaxy samples into five bins of \lgoh\ that
were chosen to have similar cumulative SFRs (except for the lowest
metallicity bin).  For each bin, we constructed an observed distribution of
luminosities using all point-sources (including HMXBs, LMXBs, and background
sources) collected for all galaxies in the bin.  These distributions are
presented as histograms for the five metallicity bins in Figure~2 ({\it black
histograms\/}).  Using our derived completeness functions, along with galactic
sky area and $M_\star$ measurements, we estimated for each galaxy the expected
contribution to the point-source luminosity distribution for unrelated
background sources from the cosmic \xray\ background (hereafter, CXB) and
LMXBs, respectively.  We utilized extragalactic number counts from
Kim \etal\ (2007) to estimate the CXB component, and the $M_\star$-normalized
LMXB relation from L19 to estimate LMXB contributions.  For all cases, the CXB
and LMXB model components (shown as green-dotted and red long-dashed curves
in Figure~2) were comfortably below the observed \xray\ point-source
distributions, and indicate that our selection criteria result in a
dominance of HMXB populations.

In Figure~3, we show the CXB-and-LMXB subtracted and completeness-corrected
cumulative XLF for each of the metallicity ranges, normalized by the cumulative
SFR in each bin.  For comparison, the average HMXB XLF model fit from L19, which consists of a
single-slope power-law with a cut-off luminosity, is displayed in each panel
({\it thin blue curves\/}).  We stress that this representation of the data was
not used when fitting for the metallicity-dependent HMXB XLF (see below), due
to its cumulative form, subtraction of model components (i.e., CXB and LMXBs),
and large corrections for completeness at the low-$L$ ends of the
distributions.  However, this representation provides guidance on the overall
form of the metallicity dependence to the shape and normalization of the HMXB
XLF.

The data in Figure~3 show clear trends with metallicity, which can be gleaned
from comparisons with the L19 HMXB XLF curve ({\it blue solid curve\/}).  HXMB XLFs at all metallicities
show similar steep power-law like slopes and SFR normalizations at $L \simlt$~(3--10)~$\times
10^{37}$~\lum.  However, at $L \simgt
10^{38}$~\lum\ the distribution flattens and extends to increasingly
higher luminosities with declining metallicity. 
For example, we find that the
number of ULXs (i.e., $L > 10^{39}$~\lum) per SFR
increases from $\approx$0.5~(\sfr)$^{-1}$ to $\approx$2~(\sfr)$^{-1}$ from
\lgoh~$\approx$9 to 8 (see Section~5.1 for further discussion on ULXs). 

The trends observed in Figure~3 indicate that the  $L_{\rm X}$--SFR--$Z$
relations that have been examined in the literature (see Section~1 and
references therein) result from a complex variation in the
bright-end of the HMXB XLF with metallicity, instead of a simple change in
normalization, as has been proposed in previous investigations (e.g., Brorby
\etal\ 2014; Basu-Zych \etal\ 2016; Ponnada \etal\ 2020).  Motivated by the
observed metallicity-dependent HMXB XLF behavior in Figure~3, we chose to model
the data using the following functional form:
\begin{equation} \begin{split}
\frac{dN_{\rm HMXB}}{dL} = & {\rm SFR} \; A_{\rm HMXB} \; \exp{[-L/L_{\rm
c}(Z)]} \times \\
 & \left \{ \begin{array}{lr} L^{-\gamma_1}  & \;\;\;\;\;\;\;(L < L_b) \\ 
L_b^{\gamma_2(Z)-\gamma_1} L^{-\gamma_2(Z)},  & (L > L_b) \\ 
\end{array}
  \right.
\end{split} \end{equation}
where
\begin{equation}
\gamma_2(Z) = \gamma_{\rm 2,\odot} + \frac{d \gamma_2}{d \log Z}[ 12 +
\log({\rm O/H}) - 8.69] ,
\end{equation}
and
\begin{equation}
\log L_{\rm c}(Z) = \log L_{\rm c, \odot} + \frac{d \log L_{\rm c}}{d \log
Z}[ 12 + \log({\rm O/H}) - 8.69 ].
\end{equation}
The above model contains seven free parameters, including both {\it
metallicity-invariant parameters} like the overall normalization per SFR,
$A_{\rm HMXB}$, the broken power-law component faint-end slope, $\gamma_1$, and
break luminosity, $L_b$, as well as the {\it metallicity-dependent} bright-end
power-law slope, $\gamma_2(Z)$, and exponential cut-off luminosity, $L_{\rm
c}(Z)$.  The latter, metallicity-dependent quantities each include solar
metallicity reference values ($\gamma_{2,\odot}$ and $\log L_{c, \odot}$) and
their first derivative with respect to the logarithm of the metallicity ($d
\gamma_2/d \log Z$ and $d \log L_{c}/d \log Z$).

To fit the parameterization specified by Equations~(1)--(3), we made use of the
MCMC approach described in Sections 4.1 and 4.3 of L19.  For a given galaxy,
the \xray\ point-source luminosity distribution (equivalent to Figure~2 but for
each galaxy individually) was modeled as:
\begin{equation}
M(L) = \xi(L) \Delta \log L \left[\frac{dN_{\rm HMXB}}{d \log L} +
\frac{dN_{\rm LMXB}}{d \log L} + {\rm CXB}(L) \right],
\end{equation}
where the $\xi(L)$ is the luminosity-dependent completeness function for the
galaxy, and $dN_{\rm LMXB}/d \log L$ and CXB($L$) are the LMXB XLF and
differential number counts, respectively.  The $dN_{\rm LMXB}/d \log L$ and CXB
model components are uniquely specified for a galaxy, given its stellar mass
and sky area, respectively.  Although we expect the LMXB component to have some
metallicity dependence, we note that (1) the metallicity range associated with
the old stellar populations that harbor the LMXBs is likely to be smaller than
the gas-phase metallicity values used here since the LMXBs would have formed at earlier epochs of cosmic history;
(2) the metallicity dependence of LMXB formation is
predicted to be weaker than that of the HMXB population (see, e.g.,
Figure~2 of Fragos \etal\ 2013a); and (3) the LMXB component
is expected to be negligible for the majority of our galaxies with \lgoh~$\simlt 8.5$,
even if some plausible metallicity dependence is applied.

In our fitting procedure, an MCMC model was
quantitatively compared with the observed distributions of \xray\ point-source
luminosities, $N(L)$, and evaluated on a global basis using the $C$-statistic (Cash~1979):
\begin{equation}
C = 2 \sum_{i = 1}^{n_{\rm gal}} \left(\sum_{j=1}^{n_{\rm X}} M_{i,j} - N_{i,j}
+ N_{i,j} \ln(N_{i,j}/M_{i,j}) \right),
\end{equation}
where $n_{\rm gal} =$~\ntot\ galaxies in our sample ($i$th index), and $n_{\rm
X} = 100$ \xray\ luminosity bins ($j$th index) spanning $\log
L$(\lum)~=~35--41.7.  We note that the majority of our galaxies do not have
non-zero completeness across all 100 \xray\ luminosity bins, and bins with zero
completeness do not contribute to $C$.

\begin{table}
{\footnotesize
\begin{center}
\caption{Metallicity-Dependent HMXB XLF Model}
\begin{tabular}{lrrrr}
\hline\hline
\multicolumn{1}{l}{\sc Parameter Name} & \multicolumn{3}{c}{\sc Units} & \multicolumn{1}{c}{\sc Value} \\
\hline\hline
$A_{\rm HMXB}$ \dotfill & \multicolumn{3}{c}{(\sfr)$^{-1}$} & 1.29$^{+0.18}_{-0.19}$ \\
$\gamma_1$ \dotfill & & &  & 1.74$^{+0.04}_{-0.04}$ \\
$\log L_b$\dotfill  & \multicolumn{3}{c}{$\log$~(\lum)}  & 38.54$^{+0.12}_{-0.30}$ \\ 
$\gamma_{2,\odot}$\dotfill  & & & & 1.16$^{+0.24}_{-0.10}$ \\
$\log L_{c,\odot}$\dotfill  & \multicolumn{3}{c}{$\log$~(\lum)}  & 39.98$^{+0.34}_{-0.14}$ \\ 
$\dfrac{d \gamma_2}{d \log Z}$\dotfill  & \multicolumn{3}{c}{dex$^{-1}$}  & 1.34$^{+0.23}_{-0.79}$ \\ 
$\dfrac{d\log L_{c}}{d \log Z}$\dotfill  & \multicolumn{3}{c}{dex~dex$^{-1}$}  & 0.60$^{+0.17}_{-0.49}$ \\ 
\hline
\multicolumn{5}{c}{Goodness of Fit Evaluation} \\
\hline
\multicolumn{1}{l}{\sc Sample} & \multicolumn{1}{c}{$C$} & \multicolumn{1}{c}{$C_{\rm exp}$} & \multicolumn{1}{c}{$C_{\rm var}$} & \multicolumn{1}{c}{$P_{\rm null}$} \\
\hline
\hline
{\bf Metallicity-Averaged (All Bins)} & {\bf 398} & {\bf 360} & {\bf 581} & {\bf 0.108} \\
$12+\log$(O/H)$ =$~8.70--9.20 & 94 & 79 & 139 & 0.218 \\ 
$12+\log$(O/H)$ =$~8.55--8.70 & 75 & 68 & 109 & 0.495 \\ 
$12+\log$(O/H)$ =$~8.38--8.55 & 69 & 79 & 140 & 0.429 \\ 
$12+\log$(O/H)$ =$~8.10--8.38 & 109 & 87 & 143 & 0.060 \\ 
$12+\log$(O/H)$ =$~7.00--8.10 & 51 & 47 & 50 & 0.574 \\ 
\multicolumn{5}{c}{} \\
{\bf Universal (All Galaxies)} & {\bf 1247} & {\bf 1040} & {\bf 1867} & {\bf $<$0.001} \\
{\it Sources with $P_{\rm null} < 0.05$}\dotfill & & & & \\
            NGC0024 & 24 & 11 & 29 & 0.019 \\ 
            NGC0337 & 26 & 11 & 25 & 0.002 \\ 
            NGC0925 & 31 & 12 & 28 & $<$0.001 \\ 
            NGC5408 & 30 & 11 & 32 & $<$0.001 \\ 
            NGC5474 & 44 & 11 & 28 & $<$0.001 \\ 
        RC2A1116+51 & 8 & 1 & 9 & 0.024 \\ 
        SBS0940+544 & 9 & 1 & 8 & 0.004 \\ 
\hline
\end{tabular}
\end{center}
}
\end{table}

%
\section{Results}
%

%
%
\begin{figure*}
\figurenum{4}
\centerline{
\includegraphics[width=19cm]{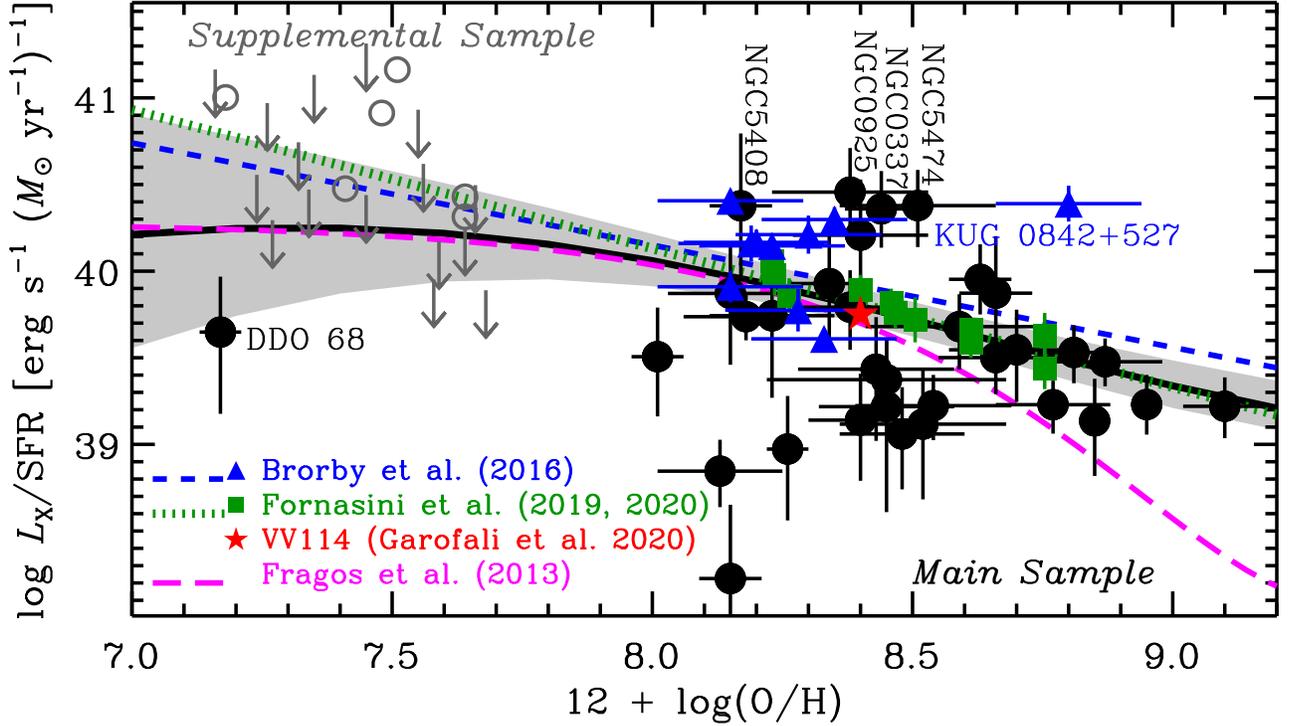}
}
\caption{
Galaxy-integrated HMXB population $L_{\rm X}$/SFR versus metallicity (\lgoh)
for the galaxies in our main ({\it black filled circles\/}) and supplemental
({\it gray open circles and 84\% confidence upper limits\/}) samples and our best-fit
metallicity-dependent XLF model ({\it black solid curve with 1$\sigma$ model
uncertainty range\/}).  Uncertainty related to stochasticity is not included in
this uncertainty range; however, its impact is discussed and quantified in
Section~5.1.  $L_{\rm X}$ values for main-sample galaxies represent the
estimated HMXB \xray\ luminosities after correcting for model contributions
from LMXB and CXB sources.  For the supplemental galaxies, $L_{\rm X}$ values
and upper limits are taken from Douna \etal\ (2015).  For comparison, we
overlay the locations of 10 Lyman break analogs from Brorby \etal\ (2016; {\it
blue triangles\/}), 12 stacked samples of $z \approx$~0.1--2.6 galaxies from
Fornasini \etal\ (2019, 2020; {\it green squares\/}), and the high-SFR,
low-metallicity galaxy VV~114 (Garofali \etal\ 2020; {\it red star\/}).  The
best-fit power-law models from Brorby \etal\ (2016) and Fornasini \etal\ (2020)
are shown as {\it blue-dashed} and {\it green-dotted} curves, respectively, and
the Fragos \etal\ (2013b) XRB population synthesis model trend is shown as a
{\it magenta long-dashed} curve. 
}
\end{figure*}

In Table~2, we provide the resulting MCMC median and 16--84\% confidence range
for each parameter in our model, which is specified in Equations~(1)--(3).  We
adopt the ``best model'' as the model in our MCMC run that produced the 
minimum value of $C$ in Eqn.~(5).  Our best model XRB source
count distributions and SFR-normalized cumulative HMXB XLFs are shown for the five
\lgoh\ bins in Figures~2 and 3, respectively.   

Our best-fit model recovers the main observed features of the
metallicity-dependent HMXB XLF discussed in Section~3, including the transition
from a nearly single power-law slope at high metallicity to a broken power-law
with a relatively flat extension at $L \approx$~(1--50)~$\times 10^{38}$~\lum\
at low metallicity.  At solar metallicity, our HMXB XLF slopes straddle the
best-fit single power-law values presented by Mineo \etal\ (2012a) and L19 and
are consistent with values presented in studies of HMXB XLFs that are based on
direct-counterpart classifications of \xray\ sources (e.g., T{\"u}llmann  \etal\
2011 and Williams \etal\ 2015 for M33 and Chandar \etal\ 2020 for M101).

We evaluated the goodness of fit of our model in both a
``metallicity-averaged'' and more stringent ``universal'' regime.  In the
metallicity-averaged regime, we assessed whether our model provided a good
characterization of the average metallicity-dependent HMXB XLF using the data
and model as represented by the five metallicity bins presented in Figure~2.
We computed the $C$-statistic following Equation~(5), replacing $n_{\rm gal}$
with $n_{\rm metal} = 5$ metallicity bins.  We then used the methods of
Kaastra~(2017) to calculate the expected $C$-statistic, $C_{\rm exp}$, and its
variance, $C_{\rm var}$, based on Poisson statistics, and evaluated the
null-hypothesis probability, $P_{\rm null}$, following:
\begin{equation}
P_{\rm null} = 1 - {\rm erf}\left( \sqrt{\frac{(C - C_{\rm exp})^2}{2 \; C_{\rm
var}}} \right).
\end{equation}
We found that in the metallicity-averaged sense, the data are statistically
consistent with being drawn from our model, with $P_{\rm null} =$~0.108
overall, and $P_{\rm null} =$~0.060--0.574 for the five bins (see Table~2).
For comparison, the L19 HMXB XLF model (with no metallicity dependence) is
ruled out, with $P_{\rm null} < 0.0001$ overall for the same data.  However,
the L19 model is acceptable only for the \lgoh~=~8.6--8.8 and 8.4--8.6 bins ($P_{\rm
null} = 0.857$ and 0.703, respectively), which is somewhat expected given that
the majority of the L19 galaxies have metallicities in these ranges.  The failure of
the L19 model for the highest metallicity and two lowest metallicity bins
further illustrates how our model is an improvement in the HMXB XLF characterization.

\begin{table*}
{\small
\begin{center}
\caption{Model-Integrated Relations with Metallicity and SFR}
\begin{tabular}{lrrrrrrrrr}
\hline\hline
 & & & & & \multicolumn{5}{c}{Median and 16--84\% $\log L_{\rm HMXB}^{\rm MC}$/SFR} \\
 & \multicolumn{4}{c}{Model Expectation} & \multicolumn{5}{c}{\rule{2.8in}{0.01in}} \\
\multicolumn{1}{c}{\lgoh} & \multicolumn{1}{c}{$N_{39}$/SFR} & \multicolumn{1}{c}{$N_{39.5}$/SFR} & \multicolumn{1}{c}{$N_{40}$/SFR} & \multicolumn{1}{c}{$\langle \log(L_{\rm X}/{\rm SFR}) \rangle$} & (SFR = 0.01)  & (0.1) &  (1) &  (10) &  (100)  \\
\multicolumn{1}{c}{(1)} & \multicolumn{1}{c}{(2)} & \multicolumn{1}{c}{(3)} & \multicolumn{1}{c}{(4)} & \multicolumn{1}{c}{(5)} & \multicolumn{1}{c}{(6)} & \multicolumn{1}{c}{(7)} & \multicolumn{1}{c}{(8)} & \multicolumn{1}{c}{(9)} & \multicolumn{1}{c}{(10)} \\
\hline\hline
7.0\dotfill & 4.29$_{-3.07}^{+7.63}$ & 1.60$_{-1.45}^{+5.64}$ & $<$1.9036 & 40.21$_{-0.66}^{+0.69}$ & 38.16$_{-0.56}^{+0.75}$ & 38.82$_{-0.50}^{+1.53}$ & 39.92$_{-0.31}^{+0.22}$ & 39.95$_{-0.08}^{+0.07}$ & 39.96$\pm$0.02 \\
7.2\dotfill & 4.27$_{-2.53}^{+5.09}$ & 1.80$_{-1.42}^{+3.76}$ & $<$1.3784 & 40.25$_{-0.50}^{+0.53}$ & 38.16$_{-0.56}^{+0.77}$ & 38.89$_{-0.56}^{+1.56}$ & 40.04$_{-0.30}^{+0.22}$ & 40.07$_{-0.08}^{+0.07}$ & 40.07$\pm$0.02 \\
7.4\dotfill & 3.97$_{-1.90}^{+3.26}$ & 1.83$_{-1.17}^{+2.38}$ & 0.20$_{-0.19}^{+0.93}$ & 40.25$_{-0.38}^{+0.40}$ & 38.16$_{-0.56}^{+0.77}$ & 38.93$_{-0.59}^{+1.59}$ & 40.11$_{-0.31}^{+0.22}$ & 40.14$_{-0.08}^{+0.07}$ & 40.14$\pm$0.02 \\
7.6\dotfill & 3.47$_{-1.33}^{+2.01}$ & 1.68$_{-0.85}^{+1.42}$ & 0.26$_{-0.21}^{+0.59}$ & 40.22$_{-0.28}^{+0.29}$ & 38.16$_{-0.56}^{+0.77}$ & 38.89$_{-0.56}^{+1.64}$ & 40.13$_{-0.33}^{+0.23}$ & 40.16$\pm$0.08 & 40.17$_{-0.03}^{+0.02}$ \\
7.8\dotfill & 2.84$_{-0.88}^{+1.19}$ & 1.41$_{-0.55}^{+0.81}$ & 0.27$_{-0.18}^{+0.34}$ & 40.16$_{-0.20}^{+0.21}$ & 38.16$_{-0.56}^{+0.74}$ & 38.83$_{-0.52}^{+1.65}$ & 40.09$_{-0.38}^{+0.26}$ & 40.14$_{-0.09}^{+0.08}$ & 40.14$\pm$0.03 \\
8.0\dotfill & 2.20$_{-0.54}^{+0.68}$ & 1.09$_{-0.33}^{+0.44}$ & 0.24$_{-0.12}^{+0.19}$ & 40.06$\pm$0.15 & 38.15$_{-0.55}^{+0.73}$ & 38.77$_{-0.46}^{+1.58}$ & 40.00$_{-0.46}^{+0.31}$ & 40.07$_{-0.11}^{+0.10}$ & 40.07$\pm$0.03 \\
8.2\dotfill & 1.62$_{-0.31}^{+0.37}$ & 0.77$_{-0.18}^{+0.23}$ & 0.19$_{-0.08}^{+0.11}$ & 39.94$_{-0.11}^{+0.12}$ & 38.15$_{-0.55}^{+0.72}$ & 38.71$_{-0.42}^{+1.37}$ & 39.85$_{-0.59}^{+0.37}$ & 39.96$_{-0.13}^{+0.12}$ & 39.97$\pm$0.04 \\
8.4\dotfill & 1.15$_{-0.18}^{+0.21}$ & 0.51$_{-0.10}^{+0.13}$ & 0.13$_{-0.05}^{+0.06}$ & 39.80$_{-0.09}^{+0.10}$ & 38.15$_{-0.55}^{+0.70}$ & 38.67$_{-0.39}^{+1.00}$ & 39.63$_{-0.67}^{+0.48}$ & 39.81$_{-0.16}^{+0.14}$ & 39.83$\pm$0.05 \\
8.6\dotfill & 0.80$_{-0.15}^{+0.17}$ & 0.32$_{-0.08}^{+0.09}$ & 0.08$_{-0.03}^{+0.04}$ & 39.64$_{-0.10}^{+0.11}$ & 38.14$_{-0.55}^{+0.69}$ & 38.65$_{-0.37}^{+0.75}$ & 39.37$_{-0.55}^{+0.57}$ & 39.64$_{-0.19}^{+0.17}$ & 39.67$\pm$0.06 \\
8.8\dotfill & 0.54$_{-0.13}^{+0.17}$ & 0.20$_{-0.06}^{+0.08}$ & 0.05$_{-0.02}^{+0.03}$ & 39.49$_{-0.12}^{+0.13}$ & 38.14$_{-0.54}^{+0.69}$ & 38.62$_{-0.35}^{+0.64}$ & 39.14$_{-0.39}^{+0.59}$ & 39.46$\pm$0.20 & 39.50$\pm$0.07 \\
9.0\dotfill & 0.37$_{-0.11}^{+0.16}$ & 0.11$_{-0.05}^{+0.07}$ & 0.03$_{-0.01}^{+0.02}$ & 39.34$_{-0.13}^{+0.15}$ & 38.14$_{-0.55}^{+0.68}$ & 38.62$_{-0.35}^{+0.58}$ & 39.01$_{-0.30}^{+0.49}$ & 39.28$_{-0.19}^{+0.22}$ & 39.33$\pm$0.07 \\
9.2\dotfill & 0.25$_{-0.09}^{+0.15}$ & 0.07$_{-0.03}^{+0.06}$ & $<$0.0181 & 39.21$_{-0.12}^{+0.16}$ & 38.14$_{-0.55}^{+0.67}$ & 38.60$_{-0.35}^{+0.55}$ & 38.95$_{-0.26}^{+0.38}$ & 39.13$_{-0.15}^{+0.20}$ & 39.19$_{-0.06}^{+0.07}$ \\
\hline
\end{tabular}
\end{center}
Note.---The modeled integrated SFR-normalized number of sources, integrated luminosity, and statistical median, 16\%, and 84\% expected values of $\log L_{\rm X}$/SFR for various SFR values.  The statistical medians and 16--84\% confidence intervals were obtained using Monte Carlo simulations described in Section~5.1, in which \xray\ populations were stochastically sampled from our best-fit model.\\
}
\end{table*}

For the universal regime, we evaluated the null hypothesis probability using the
$C$-statistic generated by Equation~(5) for all \ntot\ galaxies from the
combined main and supplemental samples.  This approach was used to determine
our best-fit parameters in Section~3 and requires that all galaxies are
reasonably well characterized by our model, since a small number of strong
outliers can have a non-negligible influence on the total value of $P_{\rm null}$.
Under this more strict evaluation, we found that our model was ruled out with
$P_{\rm null} < 0.001$ overall.  Thus, while our model provides a good
characterization of how the ``average'' HMXB XLF varies with metallicity, it
fails to provide a ``universal'' model of the HMXB XLF for all galaxies in our
sample.  

The failure of our model in the universal regime can be tied primarily to three
of the \ntot\ galaxies, NGC~925, NGC~5408, and NGC~5474, all of which are main-sample sources with
$P_{\rm null} < 0.001$ and are deemed to be significant outliers.  All four
galaxies are observed to have excesses of $\approx$2--5 times the numbers of
\xray\ point sources compared to those predicted by our model without any clear
trends with point-source luminosity (across $\log L($\lum$) =$~37--39).  In
addition to these significant outliers, there are an additional four galaxies
that have some tension with our model at the $P_{\rm null} < 0.05$ level; we
list these galaxies in Table~2.  With the exception of NGC~1569, these
galaxies also appear to have excesses of \xray\ point-sources compared to the
model. 
We expect that unmodeled physical variations in SFH, as well as gradients in the metallicities within the galaxies themselves, are plausible causes for the
deviations from the model in the outlier population; we discuss these possibilities in
more detail in Section~5.3.
Despite its non-universality, our model provides a step forward in
characterizing the metallicity dependence of the HMXB XLF, and hereafter we
make use of this model in our discussions.

Our resulting metallicity-dependent HMXB XLF can be integrated to obtain the
average $\langle L_{\rm X}$/SFR$\rangle$ versus \lgoh, a trend that has been
explicitly studied in the literature.  In Figure~4, we show $L_{\rm X}$/SFR
versus \lgoh\ for the individual galaxies that make up our main and
supplemental samples.  These values were obtained by first modeling the XLF of
a given galaxy using an HMXB component that consisted of a simple power-law
with high-luminosity cut-off plus fixed CXB and LMXB components (following the
procedure discussed in Section~3).  The value of $L_{\rm X}$/SFR for a given
galaxy was calculated by integrating the best-fit HMXB XLF for that galaxy, and
thus excludes contributions from CXB and LMXB components.

For comparison, we show $L_{\rm X}$/SFR versus \lgoh\ for (1) 10 Lyman break
analogs (LBAs) presented by Brorby \etal\ (2016); (2) 12 metallicity-binned
stacked samples of galaxies spanning the redshift range $z \approx$~0.1--2.6
from the \chandra\ Deep Field (Fornasini \etal\ 2019) and the \chandra\ COSMOS
(Fornasini \etal\ 2020) surveys; and (3) the location of the $L_{\rm
X}$(XRB)/SFR from the low metallicity (\lgoh~$\approx 8.4$) relatively distant,
but high-SFR galaxy VV~114 (SFR~$\approx 38$~\sfr), which has been spectrally
decomposed by Garofali \etal\ (2020) using \chandra, \xmm, and \nustar\
constraints. 

Using our metallicity-dependent XLF model, we calculated probabilty
distribution functions $P(\langle L_{\rm X}$/SFR$\rangle \vert$\lgoh$)$ across
a grid of \lgoh\ using our MCMC chains.  At each value of \lgoh, we calculated
the median $L_{\rm X}$/SFR and 16--84\% range implied by our model and its parameter uncertainties, which we
show in Figure~4 ({\it black curve} and {\it gray shaded region\/}).
In Col.(5) of Table~3, we tabulate the median values and 16--84\% ranges of $L_{\rm X}$/SFR
from our model.  

For comparison with past studies, in Figure~4 we overlay the best-fit power-law
models of $L_{\rm X}$/SFR versus \lgoh\ from Brorby \etal\ (2016; {\it
blue-dashed line}) and Fornasini \etal\ (2020; {\it green-dotted
line}).\footnote{We note that displayed values of $L_{\rm X}$/SFR for
comparison studies have been corrected from their original form to our adopted
IMF and \xray\ bandpass.  Specifically, Brorby \etal\ (2016) $L_{\rm X}$/SFR
values and relations have been increased by 0.25~dex to account for their use
of the Salpeter IMF, and the $L_{\rm X}$/SFR values and relation from
Fornasini \etal\ (2019, 2020) have been increased by 0.18~dex to convert their
quoted 2--10~keV luminosities to 0.5--8~keV (based on their adopted power-law
SED with $N_{\rm H} = 10^{21}$~cm$^2$ and $\Gamma = 2$).}  We further overlay
the XRB population synthesis theoretical best model median curve from Fragos
\etal\ (2013b; {\it magenta long-dashed curve\/}).  

In the \lgoh~$> 8$ regime, our metallicity-dependent HMXB XLF model exhibits a
trend in excellent agreement to the data and relation reported by Fornasini
\etal\ (2020), albeit with a somewhat divergent trend appearing below
\lgoh~$\approx 8.2$, where no data was reported by Fornasini \etal\  The Brorby
\etal\ (2016) relation is notably shallower and predicts relatively higher
values of $L_{\rm X}$/SFR in this regime, compared to those found in this study
and the Fornasini \etal\ (2020) relation.  It is possible that contributions
from LMXBs or AGN, which were not explicitly considered, could explain the
relative elevation of the Brorby \etal\ relation.  For example, the significant
outlier at \lgoh~=~8.8, KUG 0842+527, has been reported to be a radio galaxy
and have a large stellar mass (see Svoboda \etal\ 2019).  Also in this regime
we find that the Fragos \etal\ (2013b) best model appears consistent with our
model from \lgoh~=~8--8.5, but quickly diverges to much lower predictions of
$L_{\rm X}$/SFR at higher metallicities.  The divergence at metallicities above
solar is likely due to the stellar-wind prescription becoming more aggressive
in terms of removing mass from binary systems, resulting in a larger fraction
of binaries avoiding mass-transfer (and \xray\ luminous) phases.  Stellar-wind
prescriptions in this regime are highly uncertain, due to a lack of
observationally constrained mass-loss rates.

In the extreme metal-poor regime, the uncertainties on our model increase
substantially due to weak constraints.  However, the trajectory of $L_{\rm
X}$/SFR with metallicity for our model appears to flatten, diverging from a
simple power-law extrapolation from Brorby \etal\ and Fornasini \etal\  It is
certainly the case that a power-law form will become unphysical at some low
value of metallicity, since the predicted $L_{\rm X}$ diverges for the
metal-free case.  In fact, the Fragos \etal\ (2013b) model predicts that a
turnover will occur in this regime, consistent with our best fit model.
Improved constraints on the XLF in this regime, through deeper observations and
observations of additional galaxies, are required to definitively characterize
this effect.

%
%
\begin{figure*}
\figurenum{5}
\centerline{
\includegraphics[width=9cm]{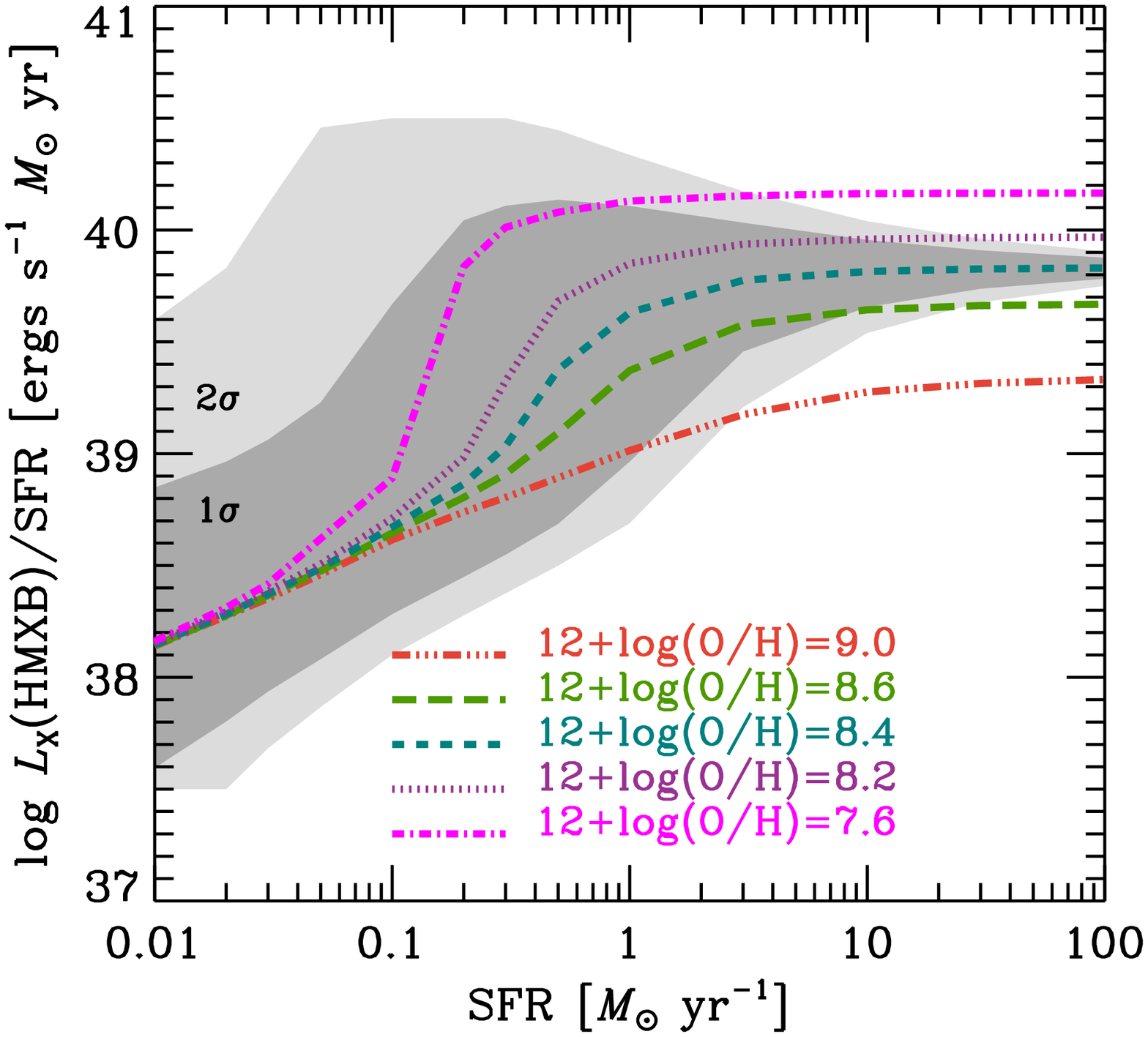}
\hfill
\includegraphics[width=9cm]{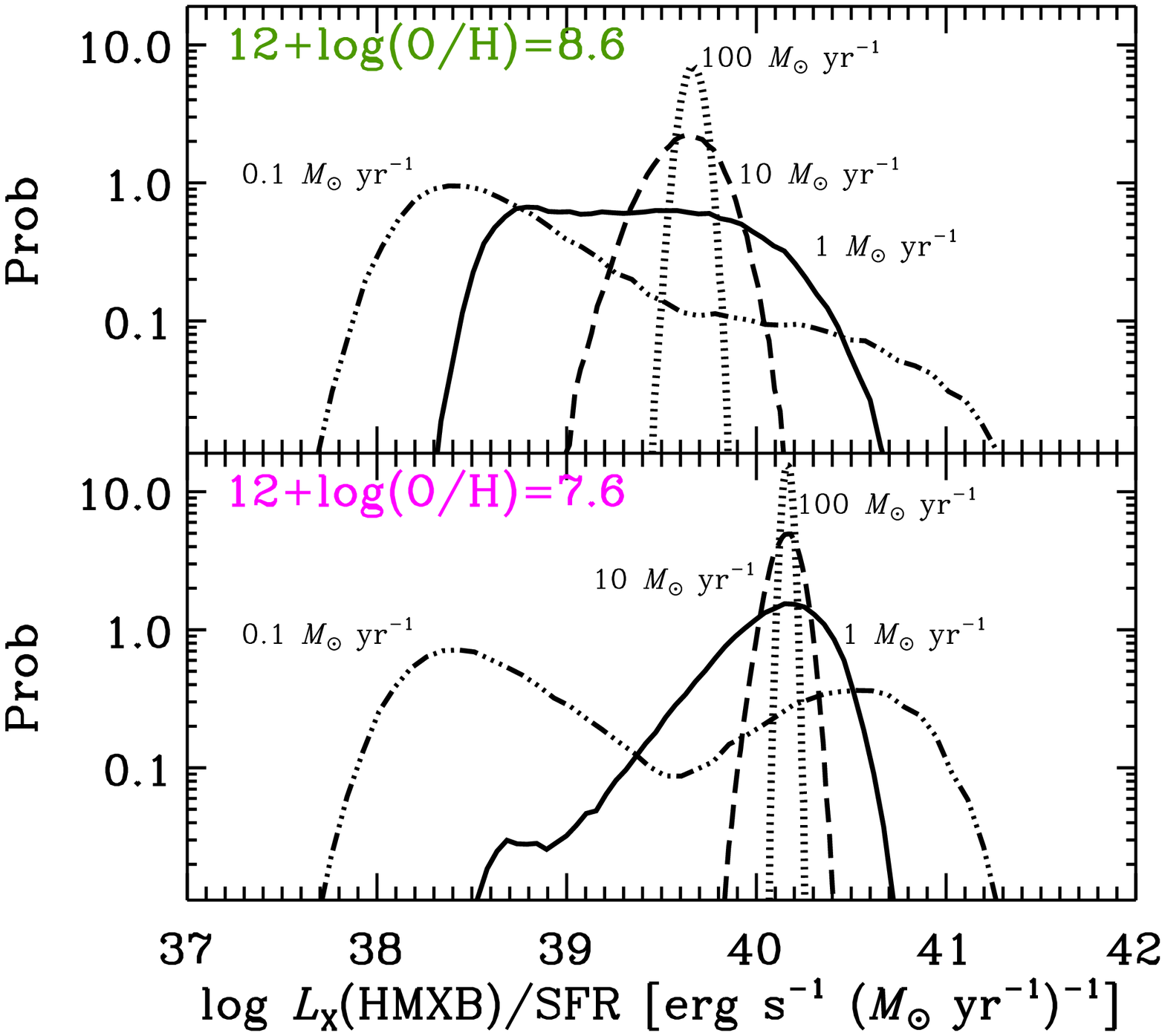}
}
\caption{
({\it Left\/}) 
Our best-fit-model predictions for the median $\log
L_{\rm X}$/SFR versus SFR for five metallicity values ({\it continuous
curves\/}; see annotation).  The predicted stochastic scatter in the
distribution of the \lgoh~=~8.4 model is shown as dark and light gray shaded
regions, representing 1$\sigma$ and 2$\sigma$ ranges, respectively.  These
regions only apply to the predicted scatter and do not include uncertainties in
model fit parameters.  The values corresponding to these curves, along with
their scatter ranges, are tabulated in Table~3, Col.~(6)--(10).  
({\it Right\/}) Example probability distributions of $\log L_{\rm X}$/SFR
associated with the stochastic scatter predicted by our best
metallicity-dependent HMXB XLF model for \lgoh~=~8.6 ({\it top}) and 7.6 ({\it
bottom}).  Similar to the left panel, these distributions include only
stochastic effects and do not contain the uncertainties associated with our
model fit parameters.  We display distributions at various SFR values (see
annotations) and two metallicities to demonstrate the variety of distributions
predicted by our model and the clear widening of the distributions going from
high-to-low SFRs.
}
\end{figure*}

%
\section{Discussion}
%

\subsection{The $L_{\rm X}$-SFR-$Z$ Relation and its Scatter}

As discussed in Section~1, the $L_{\rm X}$-SFR-$Z$ relation is of particular
importance to a variety of astrophysical investigations.  Such investigations
may require, e.g., assessments of the likelihoods that specific \xray\ detected
objects are XRBs versus AGN, or estimates of the \xray\ radiation field during
the epoch of heating at $z \simgt 10$, where HMXBs are thought to dominate.
For such studies, clear predictions for the $L_{\rm X}$-SFR-$Z$ relation and
its statistical scatter are of great use.  In the discussion that follows,
we assume that the statistical scatter is dominated by sampling stochasticity.
Additional scatter related to luminous transient and variable source
populations (e.g., BeXRBs and supergiant fast \xray\ transients; see, e.g.,
Reig~2011 and Mart{\'\i}nez-N{\'u}{\~n}ez \etal\ 2017) is expected to be
present, as well.  Thus far, such forms of scatter have not been extensively
investigated; however, a study of variable source populations in
NGC~300 finds that variability does not have a detectable impact on the
derived XLF for the galaxy (see, e.g., Binder \etal\ 2017).  Nonetheless, future studies that
assess how variable and transient source populations affect XLF scatter in a
variety of environments would help clarify this picture.

Our metallicity-dependent HMXB XLF framework can be used to quantify
statistical properties of the $L_{\rm X}$-SFR-$Z$ relation beyond the model
uncertainties discussed in Section 4.  As shown by Gilfanov \etal\ (2004), XLFs
with relatively shallow power-law slopes ($\gamma < 2$), like those
representative of the HMXB population, result in integrated $L_{\rm X}$/SFR
values that are subject to non-negligible stochastic scatter due to partial
sampling of the full-$L$ range of the XLF.  For instance, the SFR-normalized
cumulative XLF shown in the first panel of Figure~3 (i.e., for \lgoh =
8.8--9.2), suggests that a typical galaxy with SFR~=~1~\sfr\ would host a
brightest source around $L_{\rm max} \approx 3 \times 10^{38}$~\lum, while a
galaxy with SFR~=~10~\sfr\ would be expected to have $L_{\rm max} \approx
10^{40}$~\lum.  For these cases, the typical (i.e., median) values of the
$L_{\rm X}$/SFR would be dominated by the brightest sources,
\begin{equation}
L_{\rm X}/{\rm SFR} \approx \frac{1}{\rm SFR}~\int_{0}^{L_{\rm max}}
\frac{dN}{dL} L \, dL \propto \frac{L_{\rm max}^{2-\gamma}}{\rm SFR},
\end{equation}
and would take on different values for the 1~\sfr\ and 10~\sfr\ galaxy
populations.  As a corollary, populations of galaxies at a fixed metallicity
would be expected to have SFR-dependent statistical distributions and scatter
of $L_{\rm X}/{\rm SFR}$, but with a SFR-independent mean value that follows
the black curve in Figure~4.  We note that the stochastic variations of $L_{\rm
X}/{\rm SFR}$ are not expected to influence our HMXB XLFs themselves, since we
are using Poisson modeling of the differential XLF in narrow bins of $L$,
including bins that extend well beyond the most luminous detected sources.

In addition to the SFR-dependent scatter, our HMXB XLF model shape is also
metallicity dependent, which implies the statistical distributions of $L_{\rm
X}$/SFR are also metallicity dependent.
To quantify the scatter-related
SFR and metallicity dependent probability distribution functions of $L_{\rm
X}$/SFR, we employed
Monte Carlo simulations following the procedures outlined in Section~5.3 of
L19.  Briefly, a given Monte Carlo trial uses a chosen pair of SFR and \lgoh\ values
to uniquely specify our model HMXB XLF (Eqns.~1--3), which provides a prediction for the
total number of HMXBs, $N^{\rm MC}_{\rm HMXB}$, in the population and their
$L$-dependent distribution.  We then probabilistically sample from that
distribution to assign luminosity values to the $N^{\rm MC}_{\rm HMXB}$
simulated HMXBs, and sum the luminosities to generate $L_{\rm HMXB}^{\rm MC}$.
This procedure is repeated 1,000 times to build up the distribution of $L_{\rm
HMXB}^{\rm MC}$ for the given SFR and \lgoh\ pair.  We note that this procedure provides stochastic distributions of $L_{\rm HMXB}^{\rm MC}$ for the best-fit model solution and does not include parameter uncertainties.  In Table~3 (Col.~6--10), we
list the statistical properties inferred from our simulations, including the
median and 16--84\% limits on $\log L_{\rm HMXB}^{\rm MC}$/SFR for various SFR
and \lgoh\ pairs.

In the left panel of Figure~5, we show the median $L_{\rm HMXB}^{\rm MC}$/SFR
versus SFR at five metallicity values (corresponding to values close to those used in Figs.~2
and 3).  At high-SFR, all curves asymptote to the metallicity-dependent mean $L_{\rm X}$/SFR
values (i.e., Col.(5) in Table~3).  However, at SFR~$\simlt$~0.1, the curves for all
metallicities converge to the similar values.  This occurs when the expected
brightest source in the distribution is below the break luminosity ($L_{\rm
max} \ll L_{b}$), where our XLF model is metallicity independent (see the $L
\le L_b$ branch in Eqn.~1).  For reference, we also display the 1$\sigma$ and
2$\sigma$ scatter regimes for the \lgoh~=~8.4 curve.  Not surprisingly,
the scatter decreases in size with increasing SFR, varying from
$\approx$0.6~dex at SFR~=~0.01~\sfr\ to $\approx$0.05~dex at SFR~=~100~\sfr; at
the high-SFR end, the scatter is expected to smaller than the current model
uncertainty on $\langle L_{\rm X}$/SFR$\rangle$ itself (see Col.~5 of
Table~3).

In the right panel of Figure~5, we display the probability distributions of
$L_{\rm HMXB}^{\rm MC}$/SFR at four SFR values for \lgoh~$=7.6$ (bottom) and 8.6 (top).  These
distributions vary from high levels of skewness at low SFR (e.g., SFR = 0.1 and
1~\sfr) and approach Gaussian-like distributions, centered around the mean, at
SFR~$\simgt$~2--5~\sfr.  It is also apparent from the right-hand panels of Figure~5 that metallicity has a significant
impact on the shapes of the distributions.  For example, for SFR~=~0.1~\sfr,
the median values of $L_{\rm HMXB}^{\rm MC}$/SFR are similar for both
\lgoh~=~7.6 and 8.6, but their distributions are notably different at high
$L_{\rm HMXB}^{\rm MC}$/SFR.

%
%
\begin{figure}
\figurenum{6}
\centerline{
\includegraphics[width=9cm]{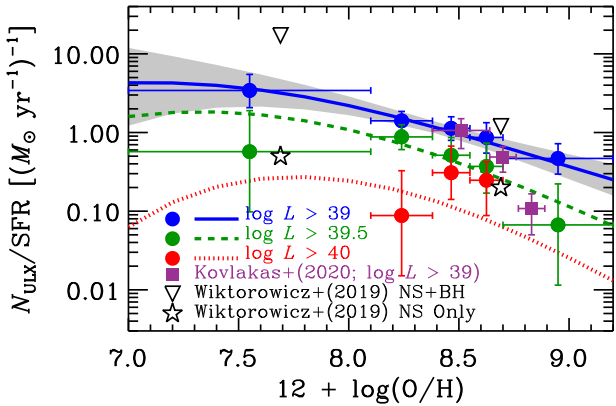}
}
\caption{
Number of ULXs per unit SFR as a function of metallicity for ULXs with $\log L
>$~39, 39.5, and 40.  Data points with full metallicity ranges and 1$\sigma$
errors on $N_{\rm ULX}$/SFR represent the observed values for our sample, and
continuous curves represent the predictions from our best-fit
metallicity-dependent HMXB XLF model.  The shaded region represents the
16--84\% confidence range for our best-fit model prediction for sources with
$\log L >$~39.  These curves (as well as their uncertainties) are tabulated in
Table~3.  For comparison, the {\it lavender filled squares} show estimates for
the frequency of $\log L$(\lum)~$> 39$ sources in 40 late-type galaxies with
$D<40$~Mpc from Kovlakas \etal\ (2020).  We also show the population synthesis
predictions from Wiktorowicz \etal\ (2019) for the expected ``observed''
$N_{\rm ULX}$/SFR with $\log L >$~39 at 0.1~$Z_{\odot}$ and $Z_\odot$ for
neutron star accretors ({\it open stars\/}) and neutron star plus black hole
accretors ({\it open downward-pointing triangles\/}).
}
\end{figure}

\subsection{Ultraluminous X-ray Source Populations}

As discussed above, our metallicity-dependent HMXB XLF analysis indicates that
the $L_{\rm X}$-SFR-$Z$ relation is driven primarily by a metallicity
dependence in the $L > 10^{38}$~\lum\ HMXB population.  Of particular interest
are ULXs (see, e.g., Kaaret \etal\ 2017b for a review), which are typically
defined as off-nuclear \xray\ sources with $L \simgt 10^{39}$~\lum.  Under
isotropic emission assumptions (as used here), ULX luminosities can exceed the
Eddington rates of 10--100~$M_\odot$ black holes.  This fact has led to an
assumption that ULXs primarily host black holes, with some potentially
having ``intermediate masses'' of $10^2$--$10^4 M_\odot$, well above black hole
masses predicted by stellar evolution (e.g., Colbert \& Mushotzky~1999).

Over the last decade, \xmm\ and \nustar\ observations have revealed nearly
ubiquitous high-energy turnovers at $E \simgt 3$~keV in ULX spectra, suggesting
relatively cool Comptonization components and/or advection-dominated accretion
disks, which are expected for stellar-mass accretors with super-Eddington
mass-transfer rates and beamed emission (e.g., Gladstone \etal\ 2009; Sutton
\etal\ 2013; Middleton \& King~2016; Walton \etal\ 2018); however, some
intermediate mass black hole candidates remain (e.g., HLX-1; Farrell \etal\
2009).  Recent discoveries of pulsations in several ULXs betray the neutron
star nature of their accretors (Bachetti \etal\ 2014; F{\"u}rst \etal\ 2016;
Israel \etal\ 2017; Brightman \etal\ 2018; Carpano \etal\ 2018; Earnshaw \etal\
2018), sometimes revealing enormous beaming factors (e.g., M82 X-2 and NGC~5907
ULX1 have observed $L/L_{\rm Edd} \approx$~100--500 assuming a 2~$M_\odot$
neutron star; Bachetti \etal\ 2014; Israel \etal\ 2017).  Detailed binary
evolution models suggest that such NS ULX binaries originate from intermediate
\xray\ binaries (e.g., Misra \etal\ 2020), which is in tension with parametric
population synthesis models that predict $\approx$$M_\odot$ giant donor stars
(e.g., Wiktorowicz \etal\ 2017).

Given their high luminosities and associations with star-forming galaxies
(e.g., Swartz \etal\ 2011; Walton \etal\ 2011; Wang \etal\ 2016; Earnshaw
\etal\ 2019; Kovlakas \etal\ 2020), ULXs are often assumed to be a subset of
the HMXB population with young massive donor stars readily capable of driving
large mass-transfer rates via Roche-lobe overflow (e.g., King \etal\ 2001).
Studies aimed at isolating ULX counterparts have revealed direct
associations with massive stars (see, e.g., Motch \etal\ 2014; Heida \etal\
2019) or close spatial proximity to young luminous star clusters (e.g., Swartz
\etal\ 2009; Poutanen \etal\ 2013; Egorov \etal\ 2017; Oskinova \etal\ 2019),
providing direct evidence that many ULXs are indeed HMXBs. However, given the
complexity of ULX accretion mechanisms, along with uncertainties in the
distributions of accretor types and beaming factors in the observed population,
theoretical models predict a large range of formation frequencies of ULXs
within galaxies (i.e., number of ULXs per SFR or $M_\star$) that would be
visible as ULXs from Earth (e.g., Belczynski \etal\ 2004; Linden \etal\ 2010;
King \& Lasota~2016; Wiktorowicz \etal\ 2017, 2019; Kuranov \etal\ 2020).
Going forward, these models would benefit from observational benchmarks such as
those provided by our study.

In Figure~6, we show our constraints on the metallicity-dependent observed
number of ULXs per SFR that have apparent luminosities (i.e., under the
isotropic emission assumption) greater than 1, 3, and 10~$\times 10^{39}$~\lum.
Our XLF model constraints are overlaid as continuous curves, with the model
uncertainty indicated as a shaded region for the $10^{39}$~\lum\ case.  These
values are tabulated in Col.~(2)--(4) in Table~3.  For comparison, we show
recent constraints from Kovlakas \etal\ (2020) on the ULX frequency as a
function of \lgoh\ for 40 late-type galaxies with $D<40$~Mpc ({\it lavendar
filled squares\/}).  These estimates are based on a galaxy subsample that
excluded elliptical/lenticular galaxies, galaxies with uncertain morphological
classification in HyperLEDA ($e_t > 1.0$), objects without complete sky
coverage, and sources where SFR and $M_\star$ were calculate in their HECATE
catalog (see Kovlakas \etal\ 2020 for details).  These estimates are in good
agreement with our estimates, except at \lgoh~$\approx 8.8$, where the Kovlakas
\etal\ (2020) ULX frequency is lower than our estimate by a factor of 2--5.
Further investigations, beyond the scope of this paper, will be required to
clarify the source of this discrepancy.

We further display, in Figure~6, the Wiktorowicz \etal\ (2019) population synthesis
predictions of the ULX frequency for the case of a continuous SFH of 6~\sfr\
over a period of 100~Myr (thus tracing HMXB ULXs) at 0.1~$Z_\odot$ and
$Z_\odot$, which correspond to \lgoh~=7.69 and 8.69, respectively.
These predictions are based on their
{\ttfamily Startrack} XRB population synthesis models (Belczynski \etal\ 2002,
2008) and the mass-transfer-rate-dependent beaming prescriptions outlined in
King \etal\ (2001) and King~(2009).  As such, the models account for the XRB
populations that would appear as ULXs along a given line of sight, while
tracking the intrinsic ULX population properties, including the accretor type.
The predictions shown in Figure~6 correspond to the expected number of {\it
observed} ULXs per SFR, after including the effects of beaming (i.e., as they
would be observed externally).
We show both the neutron star accretor and total (neutron star plus black hole)
ULX predictions.  

We find that the Wiktorowicz \etal\ (2019) predictions for neutron stars are
somewhat lower than the total rates observed for all ULXs from this study.
When including black holes, the theoretical ULX frequencies exceed the
observations presented here.  In the case of the 0.1~$Z_\odot$ population, the
theoretical models are a factor of \hbox{5--10} times higher than those
observed, suggesting that the black hole prescription itself overestimates the
prevalence of these sources.  

Since XRB population synthesis model predictions are highly variable depending
on their assumptions (see, e.g., the case study by Tzanavaris \etal\ 2013), we
are unable to make additional detailed physical conclusions about the nature of
ULX accretors based on these results.  For example, the population
synthesis work by Kuranov \etal\ (2020) shows that highly magnetized neutron
star accretors can produce ULX populations that are comparable to or larger
than those predicted by Wiktorowicz \etal\ (2019), depending on the populations
averaged magnetic moment and accretion disk structure.  These models produce
ULX luminosities through the accretion columns themselves and do not require
strong beaming.  These variations in predictions highlight the need for future
population synthesis studies that use these empirical results as direct
constraints on models.

\subsection{Towards a Universal Model for XRB XLFs}

In Section~4, we showed that our metallicity-dependent HMXB XLF model provides
a statistically acceptable prescription for the average metallicity trends.
Constraints on these trends can be improved with the combination of deeper and
more numerous observations of the galaxies in the extreme metal-poor regime
(\lgoh~$< 8$) that make up our supplemental sample (see, e.g., the constraints
in this regime from Figures~2, 3, and 4).  Building on the constraints in this
regime will also be critical for clarifying the potential impact of HMXB radiation
during the epoch of heating at $z \simgt 10$, when the majority of galaxies are
expected to have abundances in this regime (see, e.g., Torrey \etal\ 2019).  

Although our model is acceptable in a metallicity-averaged sense, it has been
ruled out as universal (see Section~4).  Given our large sample of \ntot\
galaxies, we expect $\approx$3 galaxies to have $P_{\rm null} \le 0.05$;
however, none of the three galaxies listed in Table~2 with $P_{\rm null} \le 0.001$ would be
expected by statistical scatter alone.

Some {\it observational factors} that are not inherent to the HMXB populations
themselves, such as enhanced absorption due to high disk inclinations or
morphologically disturbed systems, could distort the observed XLFs and cause
our model framework to be non-universal.  However, we have eliminated the
influence of such factors in our galaxies by our selection criteria, and the
most egregious outliers in our sample (NGC~925, NGC~5408, and
NGC~5474) are all found to have an excess of observed HMXBs over the model
predictions suggesting some other factors {\it intrinsic} to the HMXB
populations are at work.

As discussed in Section~1, in addition to metallicity, SFH variations of the
$\simlt$100~Myr population can impact both the HMXB populations, as well as
estimates on SFRs.  Explicit theoretical predictions from Linden \etal\ (2010)
and Wiktorowicz \etal\ (2017, 2019) suggest that the formation efficiencies of
ULXs (e.g., number per unit stellar mass) can vary by more than an order of
magnitude within 10--100~Myr following a burst of star formation.  Furthermore,
less luminous HMXBs in the nearby Small Magellanic Clouds and M33, have also
been shown to exhibit notable formation efficiency variations on these same
timescales as the predominant XRB donor stars transition to lower masses with
increasing age (e.g., Lehmer \etal\ 2017; Garofali \etal\ 2018; Antoniou \etal\
2019).  

Given the above, bursty SFHs like those found in low-mass, low-metallicity
galaxies could easily result in variations in the HMXB XLF and the presence of
a number of outlier galaxies in our sample.  In future work, we plan to
characterize the SFHs of a large number of local galaxies, including those in
this work, and re-assess the combined influence of metallicity and SFH on the XLF.
These steps will be crucial for ultimately developing a universal model for the
XRB XLFs in galaxies.

{\bf 
%
\section{Summary}
%
}

We have demonstrated that the SFR-normalized HMXB XLF depends on gas-phase
metallicity, and we have provided a model detailing this dependence.  Some of the key results from this study are included below.

\begin{enumerate}

\item We quantify and model the HMXB XLF across the luminosity range $\log
L$(\lum)~$\approx$~36--41 for \ngal\ galaxies in the metallicity range of
\lgoh~=~7--9.2.  However, archival data for extremely metal-poor objects with
\lgoh~$\simlt 8$ contain only dwarf galaxies with relatively shallow data,
limiting our HMXB XLF inferences in this metallicity range.

\item The impact of metallicity on the SFR-normalized HMXB XLF is most
significant for HMXBs with $L \simgt 10^{38}$~\lum.  Below this limit, the XLF
shape appears to follow a nearly universal power-law distribution, with slope
$\gamma \approx 1.7$.  Above this limit, the HMXB XLF goes from being
power-law-like at nearly solar metallicities and then developing an
increasingly flattened distribution with high-$L$ cut-off as metallicity
declines (see Figure~3).

\item We provide a new parameterization of the SFR-and-metallicity dependence
of the HMXB XLF that provides a significant statistical improvement over
constant SFR-dependent HMXB XLF models that have been widely used (see
Sections~3 and 4 and comparison in Figure~3).

\item As a consequence of our results, we establish that the
population-integrated $L_{\rm X}$--SFR--$Z$ relations reported in the
literature are driven by the metallicity-dependence of the $L \simgt
10^{38}$~\lum\ population.  For \lgoh~$\simgt 8.2$, our model predicts a
similar $L_{\rm X}$--SFR--$Z$ relation to Fornasini \etal\ (2020), but differs
somewhat from that of Brorby \etal\ (2016) and the theoretical Fragos \etal\
(2013a) relations (see Figures~4 and 5, and Section~5.1).  At \lgoh~$\simlt 8$,
our best-fit relation appears to flatten going to lower metallicity; however,
our results are highly uncertain in this regime.  Nonetheless, this behavior is
similar to the theoretical predictions from Fragos \etal\ (2013a), but is
different from the empirical extrapolations of power-law parameterizations
presented in past studies (e.g., Brorby \etal\ 2016; Fornasini \etal\ 2020).
Our model predictions for the $L_{\rm X}$--SFR--$Z$ relation and its
statistical properties are provided in Table~3. 

\item Our HMXB XLF framework also allows for observational insight into the
metallicity dependence of the ULX population.  We quantify the metallicity and
$L$ dependence of the ULX population and compare with past population synthesis
models from Wiktorowicz \etal\ (2019).  Our ULX constraints are similar to the
population synthesis predictions near solar metallicity; however, the observed
ULX population is 5--10 times lower than predictions at $\approx$0.1~$Z_\odot$.
Our constraints provide new benchmarks for future population synthesis studies
and may aid in our understanding of the nature of these complex sources (see
Section~5.2).

\item Finally, we show that a number of galaxies in our sample host HMXB
populations that are significantly elevated compared to our model predictions.
These galaxies are good candidates for harboring recent SFHs that deviate from
the average young population within galaxies.  Future studies that account for
SFH will help to quantify the effect of SFH on the HXMB population (see
Section~5.3).

\end{enumerate}

\acknowledgements

We thank the anonymous referee for their helpful comments, which have
improved the quality of this paper.  We also thank Leslie Hunt for helpful
discussions regarding galaxy sample selection.  We gratefully acknowledge
financial support from the NASA Astrophysics Data Analysis Program (ADAP) grant
80NSSC20K0444 (B.D.L., R.T.E.) and \chandra\ X-ray Center (CXC) grant
GO7-18078X (B.D.L.).  T.F.~acknowledges support from the Swiss National Science
Foundation Professorship grant (project number PP00P2 176868).

We made use of the NASA/IPAC Extragalactic Database (NED), which is
operated by the Jet Propulsion Laboratory, California Institute of Technology,
under contract with NASA.
Our work includes observations made with the NASA {\it Galaxy Evolution
Explorer} (\galex). \galex\ is operated for NASA by the California Institute of
Technology under NASA contract NAS5-98034. This publication makes use of data
products from the Two Micron All Sky Survey (2MASS), which is a joint project
of the University of Massachusetts and the Infrared Processing and Analysis
Center/California Institute of Technology, funded by NASA and the National
Science Foundation (NSF). This work is based on observations made with the {\it
Spitzer Space Telescope}, obtained from the NASA/IPAC Infrared Science Archive,
both of which are operated by the Jet Propulsion Laboratory, California
Institute of Technology under a contract with NASA.
We acknowledge the use of public data from the {\it
Swift} data archive.

{\it Facilities:} {\it Chandra}, {\it Herschel}, {\it GALEX},
Sloan, {\it Spitzer}, {\it Swift}, 2MASS, DES, Pan-STARRS

\software{ACIS Extract (v2016sep22; Broos et al. 2010, 2012), MARX (v5.3.2;
Davis et al. 2012), CIAO (v4.8; Fruscione~et al. 2006), {\ttfamily xspec} (v12.9.1;
Arnaud~1996), {\ttfamily Lightning} (Eufrasio \etal\ 2017), {\ttfamily P\'EGASE} (Fioc \& Rocca-Volmerange~1997)}

%

%

\appendix

\begin{table*}
\renewcommand\thetable{A1}
{\footnotesize
\begin{center}
\caption{X-ray point-source catalog and properties}
\begin{tabular}{llccccccccc}
\hline\hline
 &  &  $\alpha_{\rm J2000}$ & $\delta_{\rm J2000}$ & $\theta$ & $N_{\rm FB}$ & $N_{\rm H}$  & & $\log F_{\rm FB}$ & $\log L_{\rm FB}$  & Location \\
 \multicolumn{1}{c}{\sc Galaxy} & \multicolumn{1}{c}{\sc ID} &  (deg) & (deg) & (arcmin) & (counts) & ($10^{22}$~cm$^{-2}$) & $\Gamma$ & (\flux) & (\lum)  & Flag \\
 \multicolumn{1}{c}{(1)} & \multicolumn{1}{c}{(2)} & (3) & (4) & (5) & (6)--(7) & (8)--(9) & (10)--(11)  & (12) & (13) & (14) \\
\hline\hline
NGC0024 & 1 & 00 09 44.06 & $-$24 58 16.38 & 2.9 & 27.1$\pm$5.4 & 0.021 & 1.7 &  $-$14.2 & 37.6& 3 \\
  & 2 & 00 09 44.73 & $-$24 59 03.40 & 3.0 & 15.8$\pm$4.1 & 0.021 & 1.7 &  $-$14.5 & 37.3& 3 \\
  & 3 & 00 09 45.89 & $-$24 56 00.46 & 3.0 & 66.0$\pm$9.9 & 0.194$\pm$0.326 & 1.85$\pm$0.77 &  $-$13.9 & 37.9& 3 \\
  & 4 & 00 09 48.20 & $-$24 58 58.92 & 2.2 & 33.0$\pm$7.3 & 0.100$\pm$0.322 & 1.88$\pm$0.94 &  $-$14.2 & 37.6& 3 \\
  & 5 & 00 09 49.87 & $-$24 57 42.08 & 1.5 & 7.7$\pm$4.3 & 0.021 & 1.7 &  $-$14.8 & 37.0& 3 \\
\\
  & 6 & 00 09 50.25 & $-$25 00 02.24 & 2.7 & 5.3$\pm$2.4 & 0.021 & 1.7 &  $-$14.9 & 36.9& 3 \\
  & 7 & 00 09 51.27 & $-$24 59 28.45 & 2.1 & 9.8$\pm$4.6 & 0.021 & 1.7 &  $-$14.6 & 37.2& 3 \\
  & 8 & 00 09 53.62 & $-$24 58 32.01 & 1.0 & 36.0$\pm$7.5 & 0.226$\pm$0.484 & 1.68$\pm$0.97 &  $-$14.1 & 37.7& 1 \\
  & 9 & 00 09 54.63 & $-$24 56 57.60 & 0.9 & 23.0$\pm$6.3 & 1.657$\pm$4.077 & 0.98$\pm$1.76 &  $-$14.0 & 37.8& 3 \\
  & 10 & 00 09 54.85 & $-$24 57 58.96 & 0.4 & 6.6$\pm$2.6 & 0.021 & 1.7 &  $-$14.9 & 36.9& 1 \\
\\
  & 11 & 00 09 55.22 & $-$24 57 49.50 & 0.3 & 4.6$\pm$2.2 & 0.021 & 1.7 &  $-$15.0 & 36.8& 1 \\
  & 12 & 00 09 55.82 & $-$24 59 29.46 & 1.7 & 31.9$\pm$7.2 & 0.230$\pm$0.413 & $<$2.41 &  $-$14.3 & 37.5& 3 \\
  & 13 & 00 09 56.19 & $-$24 58 02.13 & 0.3 & 13.4$\pm$3.7 & 0.021 & 1.7 &  $-$14.6 & 37.2& 1 \\
  & 14 & 00 09 56.27 & $-$24 57 33.72 & 0.2 & 24.0$\pm$6.4 & 1.396$\pm$0.887 & $<$2.75 &  $-$14.3 & 37.5& 1 \\
  & 15 & 00 09 56.27 & $-$24 57 57.28 & 0.2 & 9.6$\pm$3.2 & 0.021 & 1.7 &  $-$14.7 & 37.1& 1 \\
\\
  & 16 & 00 09 57.31 & $-$24 57 42.01 & 0.2 & 25.4$\pm$5.1 & 0.021 & 1.7 &  $-$14.3 & 37.5& 1 \\
  & 17 & 00 09 58.90 & $-$24 56 57.15 & 1.0 & 8.5$\pm$3.0 & 0.021 & 1.7 &  $-$14.8 & 37.0& 1 \\
  & 18 & 00 10 00.94 & $-$24 57 27.96 & 1.0 & 19.1$\pm$5.9 & 0.021 & 1.7 &  $-$14.5 & 37.4& 3 \\
  & 19 & 00 10 03.29 & $-$24 57 30.24 & 1.6 & 19.5$\pm$5.9 & 0.021 & 1.7 &  $-$14.4 & 37.4& 3 \\
  & 20 & 00 10 03.50 & $-$24 55 28.14 & 2.8 & 17.0$\pm$5.7 & 0.021 & 1.7 &  $-$14.5 & 37.3& 3 \\
\\
\hline
\end{tabular}
\end{center}
Note.---The full version of this table contains 2915 sources, including all \nxps\ sources that were used in our XLF analyses (i.e., Flag=1).  An abbreviated version of the table is displayed here to illustrate its form and content.  A description of the columns is provided in the Appendix.\\
}
\end{table*}

In Table~A1, we provide the \xray\ point source catalogs, based on the analyses
presented in Section~3.  The columns include the following: Col.(1):
Name of the host galaxy. Col.(2): point-source identification number within the
galaxy. Col.(3) and (4): Right ascension and declination of the point source.
Col.(5): Offset of the point source with respect to the average aim point of
the \chandra\ observations. Col.(6) and (7): 0.5--7~keV net counts (i.e.,
background subtracted) and 1$\sigma$ errors. Col.(8)--(9) and (10)--(11):
Best-fit column density $N_{\rm H}$ and photon index $\Gamma$, respectively,
along with their respective 1$\sigma$ errors, based on spectral fits to an
absorbed power-law model ({\ttfamily TBABS$_{\rm Gal} \times$ TBABS $\times$
POW} in {\ttfamily xspec}).  For sources with small numbers of counts ($<$20
net counts), we adopted only Galactic absorption appropriate for each galaxy
and a photon index of $\Gamma = 1.7$.  Col.(12) and (13): the respective
0.5--8~keV flux and luminosity of the source. Col.(14): Flag indicating the
location of the source within the galaxy.  Flag=1 indicates the source has $L>
10^{35}$~\lum, is within the galactic footprint adopted in Table~1, and is
outside a central region of avoidance, if applicable.  All XLF calculations are
based on Flag=1 sources.  Flag=2 indicates that the source is within the
adopted galactic footprint, but has a luminosity of $L < 10^{35}$~\lum, and was
thus excluded from our XLF analysis. Flag=3 indicates that the source is
outside the adopted galactic footprint, but within 20\% of its outer boundary.
These sources are candidates for belonging to the galaxies, but have a high
chance of being background objects.  Flag=4 indicates that the source is
located in the central region of avoidance due to either the presence of an AGN
or very high levels of source confusion. Flag=5 indicates that the source is
$>$1.2 times the angular distance to the adopted galactic boundary. 

\end{document}